\documentclass[pra,twocolumn,a4paper, accepted=2017-10-31]{quantumarticle}  
\usepackage{bbm, amsmath,dsfont, amssymb, amsthm, bm, textcomp, nicefrac,natbib,hyperref}
\usepackage{amsthm,amssymb,amsmath,graphicx,epsfig,color,verbatim,enumerate,amsthm}
\usepackage[T1]{fontenc}
\usepackage[utf8]{inputenc}

\newcommand{\ket}[1]{\left|#1\right\rangle}
\newcommand{\bra}[1]{\left\langle#1\right|}
\newcommand{\braket}[2]{\left\langle #1|#2\right\rangle}
\newcommand\defn[1]{\textsl{#1}}

\newcommand\ketbra[2]{\left|#1\right\rangle\!\!\left\langle#2\right|}
\newcommand{\proj}[1]{\ket{#1}\!\!\bra{#1}}
\newcommand{\mean}[1]{\langle #1 \rangle}
\newcommand\cH{{\mathcal H}}
\newcommand\cM{{\mathcal M}}
\newcommand\cN{{\mathcal N}}

\newcommand\cG{{\mathcal G}}
\newcommand\cD{{\mathcal D}}
\newcommand\cI{{\mathcal I}}
\newcommand\cB{{\mathcal B}}
\newcommand\cE{{\mathcal E}}

\newcommand\cP{{\mathcal P}}%

\newcommand{\eins}{\mathbbm{1}}

\newcommand{\be}{\begin{equation}}
\newcommand{\ee}{\end{equation}}
\newcommand{\bea}{\begin{align}}
\newcommand{\eea}{\end{align}}

\def\tr{\mathrm{tr}}
\def\d{\mathrm{d}}

\newtheorem*{result*}{Result}
\newcommand\ii{\mathrm{i}}

\newcommand\blfootnote[1]{%
  \begingroup
  \renewcommand\thefootnote{}\footnote{#1}%
  \addtocounter{footnote}{-1}%
  \endgroup
}

\begin{document}
\date{\today}
\title{Macroscopic superpositions require tremendous measurement devices}
\author{M.~Skotiniotis$^\dagger$}
\affiliation{F\'isica Te\`orica: Informaci\'o i Fen\`omens Qu\`antics, Departament de F\'isica, Universitat Aut\`onoma de Barcelona, 08193 Bellatera (Barcelona) Spain}
\author{W.~D\"{u}r}
\author{P.~Sekatski$^\dagger$}
\blfootnote{$^\dagger$ These authors contributed equally to this work}
\affiliation{Institut f\"ur Theoretische Physik, Universit\"at Innsbruck, Technikerstr. 21a, A-6020 Innsbruck,  Austria}

\begin{abstract}
We consider fundamental limits on the detectable size of macroscopic quantum superpositions. We argue that a full quantum mechanical treatment of system plus measurement device is required, and that a (classical) reference frame for phase or direction needs to be established to certify the quantum state. When taking the size of such a classical reference frame into account, we show that to reliably distinguish a quantum superposition state from an incoherent mixture requires a measurement device that is quadratically bigger than the superposition state. Whereas for moderate system sizes such as generated in previous experiments this is not a stringent restriction, for macroscopic superpositions of the size of a cat the required effort quickly becomes intractable, requiring measurement devices of the size of the Earth. We illustrate our results using macroscopic superposition states of photons, spins, and position. Finally, we also show how this limitation can be circumvented by dealing with superpositions in relative degrees of freedom.
\end{abstract}
\maketitle
\section{Introduction}
\label{sec:intro}

While natural sciences exist in some form or another for more than two millennia, quantum mechanics was only formulated about a century ago. It took about fifty years to go from the basic formulation of quantum mechanics to its most radical philosophical implication---the Bell theorem~\cite{Bell64}. The reason both things took so long is quite simple: the world we perceive as human beings, outside our laboratories, is classical. There is essentially nothing quantum about it.

Today we know that there is nothing surprising about the fact that quantum effects are hidden from our everyday experience. Unavoidable interaction with the environment (decoherence) makes things look classical~\cite{Zurek03}. There is also growing evidence that the usage of imperfect detectors (such as our sense-organs) often forbids the observation of quantum phenomena~\cite{Mermin80, Peres95, Kofler07, Sekatski14a}. In fact, whenever a superposition state is deemed more macroscopic (accordingly to some definition) than another, then it is also more fragile with respect to one of the above, see e.g.,~\cite{Sekatski14b}. Nevertheless, decoherence and detector resolution are technical problems that can be overcome by working harder and harder in the lab, where we do observe quantum phenomena and violation of Bell inequalities for bigger and bigger systems~\cite{Haroche13, Ourjoumtsev07, Demartini08, Vlastakis13, Lvovsky13, Jeong14, Julsgaard01, Christensen14, Tiranov16, Arndt99, Gerlich07, Eibenberger13}.  But how far can we potentially go? Is there hope to observe a macroscopic superposition of an everyday classical object in some future lab?

Here we attempt to answer this question, and provide fundamental limits on the resources required to certify a macroscopic quantum superposition state. We concentrate on the measurement, i.e., the certification of a macroscopic quantum superposition state, although similar arguments may be used for the preparation process of such states as well.  Our approach is based on the observation that the description of any physical system in quantum mechanics in terms of a pure quantum state implicitly assumes the existence of a suitable, ideal background reference frame (RF). In reality such an RF is itself a quantum system of some limited size and can only encode a finite amount of ``unspeakable" information~\cite{Peres02}. Incorporating the RF into the quantum formalism the joint system of macroscopic superposition and RF can be treated as a closed quantum system whose physical laws are invariant with respect to fundamental symmetries of nature.  We exploit this invariance to show that, within some simple model, in order to observe a macroscopic superposition of the size of a cat one requires an RF that is of the size of the whole Earth. However, this restriction on the size of the RF can be circumvented in the case of macroscopic superpositions  of relative degrees of freedom, for which no fundamental conservation laws exist.
\section{Background and Motivation}
\label{sec:back&moti}

We begin by introducing the macroscopic quantum states that will occupy us for the remainder of this work in Secs.~\ref{sec:background} and ~\ref{sec:states}, and in Sec.~\ref{sec:task} we formally introduce the task of discriminating a macroscopic superposition from the corresponding mixture.  Next, we consider constraints on measurements imposed by the symmetry of physical laws and exemplify how such constraints arise and are circumvented in a familiar experimental situation from quantum optics~(Sec.~\ref{sec:example}).  The general treatment of measurements under symmetry, as well as how the constraints imposed by symmetries can be relaxed by appropriate systems that act as suitable RFs, is given in Sec.~\ref{sec:WAY}.  Finally, we provide a working definition of the measurement devices used to certify macroscopic quantum states in Sec.~\ref{sec:classical RFs}.
\subsection{Macroscopic quantum superpositions}
\label{sec:background}

A definitive test of macroscopic quantum behaviour is to demonstrate the trade-mark signature of quantum mechanics---quantum interference on the macroscopic scale~\cite{Arndt99, Gerlich07, Eibenberger13, Frowis17}.  This amounts to being able to distinguish between coherent and incoherent superpositions of macroscopically distinct states~\cite{Leggett80, Leggett02} and forms the basis for the remainder of this work.  Specifically, our goal is to distinguish the macroscopic coherent superposition 
\be
\ket{\Psi}=\frac{1}{\sqrt{2}}\ket{\psi_0}+\frac{1}{\sqrt{2}}\ket{\psi_1},
\label{eq:macroscopicsuperposition}
\ee
from the corresponding incoherent mixture 
\be
\Psi=\frac{1}{2}\proj{\psi_0}+\frac{1}{2}\proj{\psi_1}.
\label{eq:mixture}
\ee
Here $\ket{\psi_{0(1)}}$ are macroscopically distinct eigenstates of a classical observable, or narrow superpositions of neighbouring eigenstates. By macroscopically distinct we mean that the difference between the corresponding eigenvalues of $\ket{\psi_{0(1)}}$ is large.  In this work we shall consider the following three types of macroscopic superposition states:
\begin{description}
\item[Total photon number (or energy)]: the relevant macroscopic superposition can be taken to be 
\be
\ket{\Psi}=\frac{1}{\sqrt{2}}\left(\ket{0} + \ket{N}\right),
\label{eq:photonmacro}
\ee
where $\ket{0}$ and $\ket{N}$ are  single-mode Fock states that are eigenstates of the \defn{photon number operator} $\hat{N}$ with corresponding eigenvalues $0$ and $N$. 
\item[Total spin]: the macroscopic superposition state 
\be
\ket{\Psi}=\frac{1}{\sqrt{2}}\left(\ket{\uparrow}^{\otimes N} + \ket{\downarrow}^{\otimes N}\right),
\label{eq:spinmacro}
\ee
is the GHZ state of $N$ spin-$\frac{1}{2}$ particles (or qubits), where $\ket{\uparrow}$, $\ket{\downarrow}$ are eigenstates of $\sigma_z$ with eigenvalues $\pm\frac{1}{2}$ respectively.
\item[Delocalized object]: the superposition state in question can be taken to be
\be
\ket{\Psi}=\frac{1}{\sqrt{2}}\left(\ket{-\frac{L}{2}} +\ket{\frac{L}{2}}\right),
\label{eq:positionmacro}
\ee
i.e., a superposition of position eigenstates $\hat{X} \ket{\pm\frac{L}{2}} = \pm \frac{L}{2}\ket{\pm\frac{L}{2}}$ for a particle of mass $m$. This can be a quasi-particle describing the center of mass of a composite system. The states $\ket{\pm \frac{L}{2}}$ need not be eigenstates of the position operation but can be, for instance, Gaussian wave packets of width $\sigma$. However, for our purposes the width of an individual wave-packet $\sigma$ can be ignored as long as it is much smaller that the separation, $\sigma\ll L$.
\end{description}

\subsection{Macroscopic quantum states}
\label{sec:states}

More generally, one is also interested in states that do not have a two-component superposition structure of Eq.~\eqref{eq:macroscopicsuperposition}. Such states 
\be\label{eq:state gen}
\ket{\Psi} = \sum_n \psi_n \ket{n},
\ee
with $\ket{n}$ being eigenstates of photon number, total spin component or position of a particle, 
can also exhibit macroscopic quantum coherence $\psi_n \psi_m \ketbra{n}{m}$ between far away components $|n-m|\approx N$ if the variance of $\psi_n^2$ is large enough~\footnote{Without loss of generality we assume the coefficients $\psi_n$ to be real.}. Notice that a large variance and the presence of far-away coherence is a feature that renders such states macroscopic according to different measures, see~\cite{Frowis17}.  In order to witness such macroscopic coherences one has to be able to distinguish $\ket{\Psi}$ from any semi-classical state $\Psi$ that does not contain such coherences.  An important class of semi-classical states is given by
\be\label{eq:state gen sc}
\Psi = \sum_{n,m} \psi_n \psi_m c_{|n-m|} \ketbra{n}{m}, 
\ee
with $\lim_{N\to \infty} c_{N^{\beta'}} = 0$ for all $\beta'>\beta$. Such states do not contain coherences on the scale $|n-m| > N^{\beta}$.  Hence if there is no 
measurement that allows to distinguish $\ket{\Psi}$ from $\Psi$, there is also no way to detect coherences on the scale larger then $N^{\beta}$.  Note that for $\beta\leq1/2$ the state is typically not considered to be macroscopic quantum by many measures~\cite{Frowis17}.

It turns out that the requirements on the measuring device in order to witness macroscopic coherence are the same whether one considers the macroscopic superpositions of Eq.~\eqref{eq:macroscopicsuperposition}, or the more general macroscopic quantum states of Eq.~\eqref{eq:state gen}.  However, in the interest of delivering the main message of our work as clearly---and devoid of cumbersome technicalities---as possible we will first concentrate on macroscopic superpositions, and provide the details for the more general macroscopic states in Sec.~\ref{sec:states nogo}. 

\subsection{The Task}
\label{sec:task}

Our goal is to determine how the size of a macroscopic superposition state determines the corresponding size of the classical measuring device required to distinguish such states from their respective incoherent mixtures.  This problem corresponds to the following quantum hypothesis testing task~\cite{Holevo73,Helstrom76,Bae15}.  An experimentalist is given a \defn{single} system known to be in one of two possible states with equal prior probability~\footnote{We note that this represents the worst case scenario.};  the superposition state, Eq.~\eqref{eq:macroscopicsuperposition}, or the state of Eq.~\eqref{eq:mixture}. The experimentalist's goal is to discriminate between the two states with the maximum probability of success, given by  $P=\frac{1}{2}+t$, where
\be
t\equiv\frac{1}{2}\max_{\{M\}}\mathrm{Tr}\left[M(\proj{\Psi}-\Psi)\right]=\frac{1}{2}\left|\left|\, \proj{\Psi}-\Psi\,\right|\right|
\label{eq:tracedistance}
\ee  
is the trace distance between the states of Eqs.~(\ref{eq:macroscopicsuperposition},~\ref{eq:mixture}), and the maximization is over all possible measurements.  Formally speaking any resolution of the identity operator constitutes a valid measurement in quantum mechanics.  However, measurement is a physical process performed by a physical system that should, in principle, also be treated quantum mechanically.  

In practice one is usually given a device that prepares the hypothetical state rather then a single copy of it. Hence one is able to subsequently repeat the preparation and the measurement $n$ times in order to confirm if the device is preparing a superposition or a mixture~\footnote{We assume that the RF resets to its initial state between each run, or that one is provided with a fresh RF for each measurement. In reality the repeated use of a single RF will progressively degrade its state. The study of this degradation is beyond the scope of this paper.}.  In the limit of asymptotically many repetitions the probability to obtain a wrong conclusion after $n$ repetitions is known to follow the Chernoff bound~\cite{Chernoff:52} 
\be
P_\text{Err}^{(n)} \sim e^{-n \xi_\text{CB}}
\ee
where $\xi_\text{CB} = -\log \left(\min_{0\leq s\leq 1}\sum_i p_\Psi(i)^s p_{\ket{\Psi}}(i)^{(1-s)}  \right)$ and $p_\Psi(i)$ and $p_{\ket{\Psi}}(i)$ denote the conditional probabilities of obtaining outcome $i$ in a single measurement for the mixture and the superposition state respectively. This quantity can be related to the trace distance via the quantum Chernoff bound~\cite{QChernoff:06}
\be\label{eq:Chernoff}
P_\text{Err}^{(n)} \gtrsim(1-t)^n.
\ee
Hence, the trace distance of Eq.~\eqref{eq:tracedistance} is also directly useful for the case where one repeats the measurement many times, and allows us to establish the relation between the size of the superpositions, the size of the measurement device, and the required number of repetitions. Note that the quantum Chernoff bound can also be used to obtain an upper bound on the error $P_\text{Err}^{(n)} \lesssim(1-t^2)^{n/2}$. However, this upper bound may require collective measurements on the $n$ copies of the states.
\subsection{Experimental example}
\label{sec:example}

In order to illustrate how the notion of a RF enters into the description of a measuring device we first show how such a RF is used in a simple example from quantum optics.  This example will serve as a starting point from which we will provide our working definition of what we mean by a classical measuring device which we will explicitly state in Sec.~\ref{sec:classical RFs} 

Suppose that a two level atom is prepared in the state $\ket{\psi}=\frac{1}{\sqrt{2}}\left(\ket{0}+\ket{1}\right)$, and we wish to perform a measurement in the $x$-basis, $\{\ket{\pm}=\frac{1}{\sqrt{2}}\left(\ket{0}\pm\ket{1}\right)\}$.  What we have at our disposal are photon counting measurements, i.e., $z$-basis measurements. Thus, in order to perform an $x$-basis measurement we first need to apply a $\pi/4$-rotation about the $y$-axis to the state of the two-level atom followed by $z$-basis measurement.    

Such a rotation can be achieved via the interaction of the atom with a laser pulse in a coherent state $\ket{\alpha}$. In the rotating wave approximation, the interaction is described by the Jaynes-Cummings Hamiltonian~\cite{Scully97} and results in the unitary transformation
\be
U = e^{\gamma (\sigma_+ a - \sigma_- a^\dag)},
\label{eq:JC1}
\ee
where $\sigma_{\pm}$ are the raising/lowering operators acting on the two-level atom, $a,\,a^\dagger$ are the usual annihilation/creation operators of the field, and $\gamma$ is the interaction strength. It is important to note here that $U$ preserves the \defn{total energy} of atom-plus-field, and that in order to perform the required rotation $\gamma\approx \frac{\pi}{4 \alpha}$.  

Remarkably, an $x$-measurement is never perfect because the underlying unitary is never a perfect Rabi rotation about the $y$-axis and consequently always leaves some residual entanglement between the atom and the laser. As we show in App.~\ref{app:measurement}, the error we make in implementing the ideal rotation is equivalent to performing a noisy $z$-measurement, $\{E_0, E_1\}$,  given by 
\begin{align}\nonumber
E_0 &= \bra{\alpha} U^\dag \left(\proj{0}\otimes\eins_{a}\right) U \ket{\alpha} \nonumber \\ 
&=\left(
\begin{array}{cc}
 1-\frac{(-2+\pi )^2}{64 \pi ^2 \alpha ^2} & \frac{4-\pi }{16 \pi ^2 \alpha ^2} \\
 \frac{4-\pi }{16 \pi ^2 \alpha ^2} & \frac{\frac{1}{4}+\frac{1}{\pi ^2}+\frac{1}{\pi
   }}{16 \alpha ^2} \\
\end{array}
\right)
\label{eq:noisymeasurement1}
\end{align}
and $E_1 = \eins-E_0$.  In the limit $|\alpha|^2\gg1$, the average overlap of $\{E_0, E_1\}$ with the ideal $z$-basis measurement is $\bar{F}=1-\frac{0.22}{|\alpha|^2}$.  One can treat the general case of a spin-$N/2$ system with $N+1$ energy eigenstates $\ket{N/2,m}$, that are eigenstates of $J_z$ in a similar fashion (see App.~\ref{app:measurement}). The quality of the overall measurement, as quantified by the average overlap of each element with the ideal one $\bar F = \frac{1}{N+1}\sum_{m=0}^N \tr (E_m \proj{N/2,m}_\text{x})$, is
\be\label{eq:spin N}
\bar{F}=1 -0.072\frac{N(N+2)}{|\alpha|^2}+O\left(\frac{N^3}{|\alpha|^3}\right).
\ee

The coherent state $\ket{\alpha}$ serves as a RF for phase, relative to which the requisite rotation $U$ can be applied to the state of the system before performing the $z$-basis measurement.  By tuning the phase of the coherent state, as well as the interaction strength $\gamma$, we can realise all possible measurements that can be performed on the state of the atom. Note that the quality of the measurement in  Eq.~\eqref{eq:spin N} depends on the ratio between the \emph{square} of the corresponding spin-system~\footnote{Here we use the Schwinger representation to map any arbitrary $N$-photon state to the state of a spin-$N/2$ particle}, $N^2$, and the number of photons in the laser pulse $|\alpha|^2$. This is a typical relation that we will encounter throughout the remainder of this work. 
\subsection{Measurements under symmetry}
\label{sec:WAY}

\begin{figure}[hbt]
\includegraphics[width=1\columnwidth]{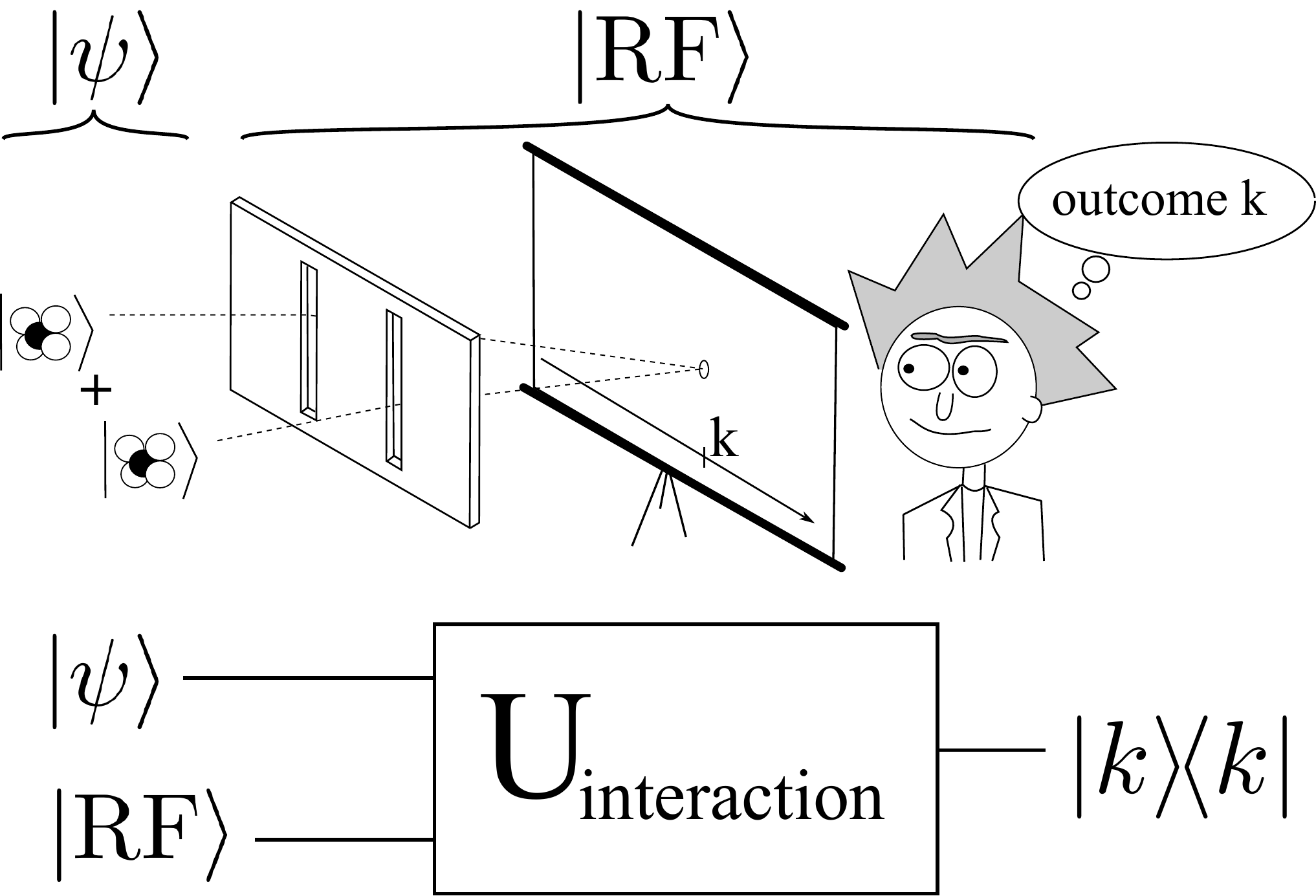}
\caption{Measuring process under symmetry.  The measurement device is described by the state $\ket{\text{RF}}$.  The measurement process consists of interacting the device with the system of interest via a globally invariant unitary interaction $U_{\mathrm{inter}}$ followed by a globally invariant von Neumann measurement $\proj{k}$.}
\label{fig:detection}
\end{figure}

The previous example illustrates how the size of the reference limits the quality of a measurement in a particular setup. However, the same argument can be made more generally. Any measurement device is comprised of a state, $\ket{\text{RF}}$, that encodes all pertinent information regarding the device's size, orientation, {\it etc.}, and a measurement process consisting of an interaction, $U$, between the state of the system, $\ket{\psi}$, and the state of device $\ket{\text{RF}}$, followed by the von Neumann measurement $\proj{k}$ on some part of the global system that realizes outcome $k$, see Fig.~\ref{fig:detection}.
The probability to observe the latter is given by
\be
p_k = \mathrm{Tr}\left[(\proj{k}\otimes\eins) \, U \proj{\psi}\otimes\proj{\text{RF}} U^\dag\right].
\label{eq:probability1}
\ee

Importantly, by explicitly incorporating the state of the RF (the measurement device) within our quantum mechanical description, the joint system-plus-device can be treated as a closed system, and as such must respect the fundamental symmetries of natural laws.  The most general, non-relativistic, group of space-time symmetries of physical laws is the Galilean group of transformations---time translations, space translations, rotations, and boosts. Noether's theorem~\cite{Noether18} then  tells us that to each such symmetry corresponds a conservation law for the closed system: conservation of energy, momentum, angular momentum {\it etc}. In particular, the only interactions, $U$, that can be performed are those that satisfy $[U,V_g]=0,\,\forall g\in G$  where $G$ is the relevant symmetry group and $G\supset g\to V_g$ is its corresponding unitary representation. Similarly, this also restricts the possible measurements that can be preformed on the closed system after the interaction; the POVM elements that can be performed have to satisfy $[\proj{k},V_g]=0, \,\forall g\in G$~\cite{Wick52, Wigner52, Araki:60, Aharonov67, Ozawa:02}.  

The invariance of the measurement might, at first sight, appear less intuitive than that of the interaction.  After all, the measurement performed by our eyes, or other sense-organs, do not commute with the group of spatial rotations. However, in this example the retina of the eye, as well as the rest of the observer's body, encode information about the observer's spatial orientation. As such, they are included in the description $\ket{\text{RF}}$, and the assumption $[\proj{k},V_g]=0$ only holds when all physical systems that can encode the relevant information are properly accounted for in the state of the RF.  The action of the projector $\proj{k}$ ultimately stands for the mere awareness of the outcome $k$ by the observer, which is invariant under the relevant group of symmetries.

The fact that both the measurement and the interaction commute with $V_g$ ensure that the probability $p_k$ in Eq.~\eqref{eq:probability1} is left unchanged when the global state is transformed $\ket{\psi}\ket{\text{RF}}\to V_g \left(\ket{\psi}\ket{\text{RF}}\right)$.  As this holds for any $g\in G$, it also holds if $g\in G$ is taken at random according to the invariant measure, $\d g$, of the group yielding
\be
p_k = \mathrm{Tr}\left[(\proj{k}\otimes\eins) \, U \left(\cG\left[\proj{\psi}\otimes\proj{\text{RF}}\right]\right) U^\dag\right],
\label{eq:probability3}
\ee
where 
\be
\cG[\proj{\psi}\otimes\proj{\text{RF}}] = \int \mathrm{d}g\, V_g\left(\proj{\psi}\otimes\proj{\text{RF}}\right)V_g^\dag
\label{eq:G-twirling}
\ee
is called the \defn{G-twirling map}. Eq.~\eqref{eq:probability3} shows that the restrictions imposed by symmetry on operations and measurements are \defn{operationally} indistinguishable from the situation where the joint state of system-plus-device is described by $\cG[\proj{\psi}\otimes\proj{\text{RF}}]$ instead of $\ket{\psi}\otimes\ket{\text{RF}}$.  This is precisely the state that one would ascribe to the joint system-plus-device in the absence of an additional external RF (see App.~\ref{app:G-twirling} for a more detailed explanation and~\cite{Bartlett07} for the complete theory).

It is important to note that for moderately sized systems a fully quantum treatment of the RF is not required as one can consider the latter to be sufficiently large, e.g., by increasing the power of a laser beam to obtain a proper phase reference (as shown in the example above), or consider the spatial orientation to be given by the laboratory equipment. However, this is no longer justified if one considers quantum objects of macroscopic size.  In this case, the explicit inclusion of the RF into our description is necessary in order to take into account the required resources.  
\subsection{Classical states of reference frames}
\label{sec:classical RFs}
The preceding discussion emphasizes how all pertinent information of the measuring device, such as its size, orientation, phase {\it etc.}, are described by the quantum state $\ket{\text{RF}}$. However, as the measurement device is by all accounts a classical system, the quantum state $\ket{\text{RF}}$ describing it must be as classical as possible.  We define states of any macroscopic quantum system, comprised of $N$ constituent parts, as \emph{classical} if their collective quantum effects can be explained as the accumulation of $\mathcal{O}(N)$ microscopic quantum effects.  This definition includes the so-called pointer states~\cite{Zurek03, Eisert:04}, and generalized coherent states~\cite{Boixo07}. In addition such states are known to be robust fixed point of generic \emph{local} decoherence processes~\cite{Zurek93, Boixo07}.  
Finally, one can also show that after the action of generic local noise  any state  can be decomposed as a mixture of pure states with the average variance of any extensive quantity that only scales linearly with the number of subsystems.  Note also that the minimum average variance of a state is equal to its quantum Fisher information~\cite{Yu13}---a measure of the states metrological performance.
Thus, for the remainder of this work, classical measuring devices will be described either by coherent states of light, spin-coherent states, or products of localized wave packets, corresponding to classical measuring devices for energy, angular momentum, and position respectively. However, in Sec.~\ref{sec:states nogo} we will show that our results can be formulated without recourse to an explicit state for a RF: all we need to assume is how its average variance scales with respect to its size.

\section{Detecting Macroscopic Superpositions}
\label{sec:Results}

In this section we derive the size of the classical measuring device needed in order to discriminate between a macroscopic superposition and a classical mixture in a single shot as explained in Sec.~\ref{sec:task}.  Specifically, we will show that the macroscopic superposition states given in Eqs.~(\ref{eq:photonmacro},~\ref{eq:spinmacro},~\ref{eq:positionmacro}) can be distinguished from their corresponding classical counterparts with an exponentially small probability of error if the classical measuring devices used are quadratically larger in size.  In Sec.~\ref{sec:states nogo} we show that the same result holds for the more general macroscopic quantum states of Eq.~\eqref{eq:state gen}.  In order to provide as much physical insight into the results as possible, we give a heuristic account of our findings below, with the rigorous mathematical proof deferred to Apps.~\ref{app:results} and~\ref{app:macro_states}.

\subsection{Macroscopic Superpositions of photon number}
\label{sec:photonmacro}

We begin with the case of distinguishing the macroscopic superposition of Eq.~\eqref{eq:photonmacro} from the incoherent mixture 
\be
\psi=\frac{\proj{0}+\proj{N}}{2}.
\label{eq:photonmixture}
\ee
Observe that in the absence of an appropriate phase reference Eq.~\eqref{eq:photonmacro} is operationally indistinguishable from that of Eq.~\eqref{eq:photonmixture}~\footnote{We remark that the distinguishability of the superposition of coherent states $\ket{\text{SCS}}=\frac{1}{\sqrt{2}}\left(\ket{\alpha}+\ket{-\alpha}\right)$ from the mixture $\frac{\proj{\alpha}+ \proj{-\alpha}}{2}$ does not suffer from a lack of a RF for total photon number. Instead the relevant group of transformations here is the set of Bogoliubov transformations associated with the Lorentz group of symmetries of special relativity.}; there exists no phase-invariant measurement capable of distinguishing between these two states.

It follows that in order to have any chance to distinguish between the two states we need a phase reference.  Hence, our classical measurement device is described by a coherent state of the RF with $|\alpha|^2$ photons in total
\be
\ket{\text{RF}}=\ket{\alpha}=\sum_{n=0}^\infty \sqrt{q_{n}}\ket{n},
\ee
 with $q_{n}=|\alpha|^{2n} e^{-|\alpha|^2}/n!$.  The joint system-plus-device is invariant under global phase shifts, i.e., for $\rho\in\cB(\cH_S), \: V_\theta (\proj{\text{RF}}\otimes \rho)V_\theta^\dagger=\proj{\text{RF}}\otimes \rho$, where $V_\theta=\exp(\ii\theta(\hat{N}_{\mathrm{RF}}+\hat{N}_{\mathrm{S}}))$ with $\hat{N}_{\mathrm{RF(S)}}$ the photon number operator for the RF (system) respectively.  In other words the total photon number (or equivalently the total energy) of the joint RF-plus-system is conserved.

Now recall that the constraints due to global phase shifts can be equivalently described by replacing the joint state of the RF-plus-system by its $G$-twirled version Eq.~\eqref{eq:G-twirling}.  As the latter involves averaging over all possible $\theta\in[0,2\pi)$, its effect on an arbitrary state of system-plus-device is to erase \defn{all} coherences between states with different \defn{total photon number} $\hat K\equiv \hat N_{\mathrm{RF}}+\hat N_{\mathrm{S}}$, resulting in a mixture (a direct sum) of states with different fixed total photon number (see App.~\ref{app:G-twirling}). Hence the $G$-twirling map reads
\be\label{eq: orth proj}
\cG\big[\proj{\text{RF}}\otimes \rho\big]= \bigoplus_{K=0}^\infty \underbrace{\Pi^{(K)}\big(\proj{\text{RF}}\otimes \rho\big)\, \Pi^{(K)}}_{Q_K \varrho^{(K)}}, 
\ee
with
\be
 \Pi^{(K)} =\sum_{n=0}^K\proj{n}_\text{RF}\otimes\proj{K-n}_\text{S}.
\ee

For $K<N$ the state $Q_K \varrho^{(K)}$ for both the macroscopic superposition state of Eq.~\eqref{eq:photonmacro} and the corresponding mixture of Eq.~\eqref{eq:photonmixture}, reads $Q_k \varrho^{(K)} = q_K \proj{K}_\text{RF}\otimes \proj{0}_\text{S}$. For $K\geq N$ however, one gets
\begin{align}\nonumber
\Pi^{(K)}\big(\proj{\text{RF}}&\otimes \proj{\psi} \big)\, \Pi^{(K)}\nonumber\\&=\frac{1}{2}\left(\begin{array}{cc} q_K & \sqrt{q_Kq_{K-N}}\\ \sqrt{q_Kq_{K-N}}& q_{K-N}\end{array}\right)
\end{align}
and
\begin{align}
\Pi^{(K)}\big(\proj{\text{RF}}&\otimes \psi \big)\, \Pi^{(K)}\nonumber \\&=  \frac{1}{2}\left(\begin{array}{cc} q_K & 0\\ 0& q_{K-N}\end{array}\right),
\end{align}
expressed in the basis $\{\ket{K}_\text{RF} \otimes \ket{0}_\text{S}, \ket{K-N}_\text{RF} \otimes \ket{N}_\text{S}\}$. The trace distance in Eq.~\eqref{eq:tracedistance} between the two states after the twirling is now very easy to compute and yields
\be\label{eq:tracedistancephotons}
t= \sum_{K=N}^\infty  \frac{1}{2}\sqrt{q_K q_{K+N}} \approx \frac{1}{2}e^{-\frac{N^2}{8|\alpha|^2}},
\ee
where the approximation holds in the limit of large $\alpha$ and $N$, which is the regime we are interested in.

For completeness, note that this trace distance and the corresponding optimal probability to distinguish the states is attained with the phase-invariant POVM composed of two elements 
\be
M_\pm=\bigoplus_{K=0}^\infty \proj{\pm_K},
\label{eq:optimalmeasurementphotons}
\ee
where
\be
\ket{\pm_K}=\frac{1}{\sqrt{2}}(\ket{K}_{\mathrm{RF}}\otimes\ket{0}_{\mathrm{S}}\pm\ket{K-N}_{\mathrm{RF}}\otimes\ket{N}_{\mathrm{S}}).
\label{eq:optimalmeasurementcomponetns}
\ee

Recalling that $\ket{\alpha}$ describes our measurement device, Eq.~\eqref{eq:tracedistancephotons} states that the probability of successfully distinguishing the macroscopic superposition from the incoherent mixture in a single shot approaches unity so long as our classical measurement device is quadratically larger than the size of the macroscopic superposition ($|\alpha|^2> N^2$).  
\subsection{Macroscopic Superpositions of Spin}
\label{sec:spinmacro}

Next, consider using a classical measuring device in order to distinguish between the macroscopic superposition of Eq.~\eqref{eq:spinmacro} and the incoherent mixture
\be
\psi=\frac{\proj{\uparrow}^{\otimes N} + \proj{\downarrow}^{\otimes N}}{2}.
\label{eq:su(2)mixed}
\ee
In the absence of a RF for orientation in space, Eq.~\eqref{eq:spinmacro} is operationally indistinguishable from Eq.~\eqref{eq:su(2)mixed}.  The underlying symmetry in this case is the invariance of physics under rotations $V_{\bf n} = \exp( \ii \,{\bf n}\cdot({\bf S}_\text{S} + {\bf S}_\text{RF}))$, where ${\bf S} = (S_X\, S_Y\, S_Z)$ are the total spin operators for the system and RF respectively. The resulting $G$-twirling map corresponds to averaging over all possible orientations of the joint system-plus-device and results in the elimination of all coherences between subspaces corresponding to different values of the total spin ${\bf S}^2 = ({\bf S}_\text{S} + {\bf S}_\text{RF})^2$, as well as the complete elimination of coherences between various directions within each subspace (see App.~\ref{app:G-twirling} for further details).  It follows that an orientational RF state needs to have some spread for its total spin ${\bf S}_\text{RF}^2$. A classical state of RF for orientation carried by $M$ spin-$\frac{1}{2}$ subsystems is, for example, given by $\ket{\text{RF}_{\bf n}}=\ket{\uparrow}^{\otimes M/3}\ket{\rightarrow}^{\otimes M/3}\ket{\odot}^{\otimes M/3}$, with $\ket{\uparrow}$, $\ket{\rightarrow}$ and $\ket{\odot}$ being the eigenstates of $S_X$, $S_Y$ and $S_Z$ operators for a spin-$\frac{1}{2}$. 

We will consider the problem of a full orientation RF in Sec.~\ref{sec:macrorfs}.  However, for the purposes of distinguishing the superposition state of Eq.~\eqref{eq:spinmacro} from the mixture of Eq.~\eqref{eq:su(2)mixed} a simpler argument involving only a directional RF suffices. A random rotation can be thought of as first choosing a direction $\hat{\bm n}$, followed by a random rotation by an angle $\theta\in(0,2\pi]$ about $\hat{\bm n}$.  Hence, the $G$-twirling map for an orientation RF includes the $G$-twirling map of a directional RF.  Choosing without loss of generality $\hat{\bm n}=\hat{\bm z}$, and averaging over all possible rotations about the $\hat{\bm z}$ axis results in an isomorphism between a directional RF and the phase reference RF of Sec.~\ref{sec:photonmacro}.  Projections onto the total photon number are now replaced by projecting the joint state of system-plus-RF onto sectors with a fixed value of $S_Z = S_{Z,\mathrm{S}}+S_{Z,\text{RF}}$, and the state of the corresponding classical RF, representing the magnetic field direction of the magnet in the Stern-Gerlach experiment, is given by a spin coherent state 
\be
\ket{\text{RF}_Z} =\ket{\rightarrow}^{\otimes M}=\sum_{m=-J}^{J} \sqrt{\frac{1}{2^M}\binom{M}{m+\frac{M}{2}}}\ket{J,m},
\label{eq:directionalrf}
\ee 
with $S_{Z,\text{RF}}\ket{J,m} = m \ket{J,m}$ and $J=\frac{M}{2}$. As the distribution of the spin-coherent state is well approximated by a Gaussian, we obtain for the trace distance
\be
t \approx \frac{1}{2} e^{-\frac{N^2}{8 M}}.
\ee
Again, one requires that the magnets be at least quadratically larger than the size of the macroscopic superposition in question.
\subsection{Macroscopic superpositions of position}
\label{sec:positionmacro}

Let us now consider the situation where the task is to distinguish a system of mass $m$, say an atom, in a superposition at two spatial locations separated by a distance $L$, i.e., Eq.~\eqref{eq:positionmacro} versus the incoherent mixture
\begin{equation}
\Psi=\frac{\proj{-\frac{L}{2}}+\proj{\frac{L}{2}}}{2}.
\label{eq:positionmixed}
\end{equation}
This corresponds to the case where one lacks a RF for the two remaining elements of the Galilean group; spatial translations and boosts. 

In the case of spatial translations the unitary representation is given by $V_s= e^{\ii s \sum_i p_i}$, where $p_i$ are the momenta of the particles in the closed system. Invariance under spatial translation implies that the joint state of the system-plus-device, $\varrho = \rho \otimes \proj{\mathrm{RF}}$, can be replaced by its $G$-twirled version
\be
\cG_P [\varrho]= \lim_{\omega\to\infty}\int_{0}^{\omega} ds \,e^{\ii s P} \varrho\, e^{-\ii s P},
\ee
which corresponds to a projection onto sectors with a fixed total momentum $P\equiv \sum_i p_i$. 

In the case of boosts the fundamental symmetry is the invariance of physical laws for all observers in relative motion, i.e., observers that move with a constant velocity with respect to each other. A boost at some initial time $t=0$ corresponds to a coordinate change $t'=t$ and $x'=x+v t$, so at $t=0$ an observer in $R'$ sees all the particles with shifted momenta $p_i'= p_i + m_i v$ as compared to the momenta $p_i$ of the particles observed in $R$. Hence, the transformation between the states in the two frames at this time is given by $V_v = e^{\ii v \sum_i x_i m_i}$. Because the relation between the two observers $R$ and $R'$ is explicitly time dependent the treatment of the boost in the Hamiltonian formalism, where time plays a special role, slightly differs from the rest of the symmetries of the Galilean group. In particular, the boost transformation does not commute with the Hamiltonian of a closed system,
\be\label{eq:H particles}
H = \sum_i \frac{p_i^2}{2 m_i} + V(x_1,\dots,v_N).
\ee
Instead, one finds $V_v^\dag H V_v = v \sum_i p_i \equiv v P$. This term ensures the proper relation between the coordinates $x' = x +v t$ of the two observers at all times. 

For example, consider the situation where a boost applied at time $t=0$ is later on inverted at time $t=\tau$. The overall transformation (boost+free evolution+inverted boost) is not identical with the free evolution $H$ in Eq.~\eqref{eq:H particles} as
\be\label{eq:boostedH}
V_v^\dag e^{\ii \tau H} V_v = e^{\ii \tau H + \ii v t P}\neq e^{\ii \tau H}. 
\ee
In fact, the additional term $e^{\ii v t P}$ exactly accounts for the coordinate shift $x_i' \to x_i+ v \tau$ acquired between the two frames during this time~\footnote{The requirement $V_v^\dag e^{\ii t H} V_v = e^{i v t P}$ enforces the dependence of the Hamiltonian on the momenta to be given by the well known form of the kinetic energy}. Hence, the twirling map
corresponding to the boost and generated with the center of mass position operator $X_C \equiv \frac{1}{M} \sum_i x_i m_i$ (with the total mass $M\equiv \sum_i m_i$) does not commute with the free evolution. However, considering the translation and boost symmetries together,  the twirling map for the momentum and the position of the center of mass,
\be\label{eq: GCM}
\cG[\varrho]= \int dv ds \,e^{\ii (v X_C+s P)} \varrho\, e^{-\ii (v X_C+s P),}
\ee 
does commute with the free evolution $\cG[e^{\ii t H} \varrho e^{-\ii t H}] = e^{\ii t H} \cG[ \varrho ]e^{-\ii t H}$. Indeed, one sees that $e^{\ii t H} e^{\ii (v X_C+s P)} e^{-\ii t H} = e^{\ii (v X_C+s P +v t P)}$ (up to an irrelevant phase) can be brought to the initial form $e^{\ii (v X_C+s P)}$ by a simple change of integration variables in Eq.~\eqref{eq: GCM}.

The $G$-twirling map washes out the center of mass  degree of freedom, seen as a quasi-particle respecting the canonical commutation relation $[X_C,P]=\ii$ (see e.g.~\cite{Messiah}). The twirling map performs a random displacement in the phase space of the center of mass variables $X_C$ and $P$, ensuring that this degree of freedom is totally undefined. This simply means that for a closed system (once the RF is internalized) only the relative degrees of freedom make physical sense, while the center of mass position and momentum are a mathematical fiction.  

Let us now focus on the effect of the $G$-twirling on the delocalized mass of  Eq.~\eqref{eq:positionmacro} and a suitable RF state. For the purpose of upper bounding the trace distance between the superposition and the mixture, the twirling over the total momentum can be omitted, and we focus only on the twirling of the the center of mass position, corresponding to the integration of $v$ in Eq.~\eqref{eq: GCM}. The latter projects the overall state 
\be
\cG_{X_C}[\varrho]= \bigoplus_{X_C}\, \Pi_{X_C} \varrho\, \Pi_{X_C},
\ee
onto orthogonal sectors with a fixed position for the center of mass $X_C \equiv \frac{m x + M X}{m+M}$, where $M$ and $X$ are the mass and position of the center of mass of the RF. We assume the latter to be a product state of $K$ particles of mass $m_0$, whose individual wave-functions $|\braket{x_i}{\psi_i}|^2$ are Gaussians centered around fixed positions with equal spread $\sigma_0$. Hence, the center of mass of the RF is a quasi-particle with mass $M = K m_0$, position operator $X \equiv\frac{\sum m_0  x_i}{M}=\frac{1}{K}\sum x_i$ and wave-function
\be
|\text{RF}(X)|^2=\tr \proj{X} \proj{\text{RF}} = \frac{1}{\sqrt{2 \pi /K} \sigma_0 } e^{-K  \frac{X^2}{2 \sigma_0^2}}.
\ee
One can then directly compute the trace distance of Eq.~\eqref{eq:tracedistance} between the states  $\cG_{X_C}[\proj{\Psi} \otimes \proj{\text{RF}}]$ and $\cG_{X_C}[\Psi \otimes \proj{\text{RF}}]$, which results in (see App.~\ref{app:Positional} for more details)
\begin{align}
t=\frac{1}{2} \int \text{RF}(X) \text{RF}^*(X+L\frac{m}{M})dX \nonumber\\
=\frac{1}{2} \exp\left(-\frac{\left(\frac{L}{\sigma_0}\right)^2 \left(\frac{m}{m_0}\right)^2}{8 K}\right).
\label{eq:massive size}
\end{align}

Yet again, the size of the RF, $K$, has to be quadratically larger then the product of the ratio of masses multiplied by the ratio of spreads between the superposed object and the particles constituting the RF.  
\subsection{Macroscopic quantum states}
\label{sec:states nogo}

In this section we present the extension of the results above to general states $\ket{\Psi}$ of Eq.~\eqref{eq:state gen}, as well as mixed states of the reference frame system $\rho_\text{RF}$. We give a simplified account of our findings;  details of the calculation can be found in App.~\ref{app:macro_states}.  Firstly, we need to identify a semi-classical counterpart, $\Psi$, of Eq.~\eqref{eq:state gen sc} which does not contain macroscopic coherences.  This is not as trivial as it might seem at first sight as an arbitrary choice of the ``masking'' function  $c_{|n-m|}$ does not necessarily yield a valid density matrix.  One possible choice that guarantees the positivity of $\Psi$ is to associate the latter as the outcome of a completely positive, trace-preserving map acting on the original state $\ket{\Psi}$.  In particular we 
choose   
\be
\Psi =\sum_{n,m} \psi_ n \psi_m \exp\left(-\frac{(n-m)^2}{2N^{2 \beta}}\right) \ketbra{n}{m},
\label{eq:Gaussian_channel}
\ee
corresponding to the action of a Gaussian dephasing channel $\Psi \equiv \cE_\beta(\ket{\Psi})$ on the initial state. This channel corresponds to the application of a random unitary $U_\lambda \equiv e^{\ii \lambda (\sum_n n \ketbra{n}{n})}$ on the initial state, where $\lambda$ is a random variable with associated probability distribution $p(\lambda)=\frac{N^\beta}{2\sqrt{\pi}}e^{-\frac{N^{2\beta} \lambda^2}{4}}$.  Notice that this noise process is solely considered to construct a classical state (without long-ranged coherences) that has the same diagonal elements as the state in question. It is not that we assume that this (or any other kind of) noise actually acts on states.

Accordingly, the trace distance between the state $\ket{\Psi}$ and $\Psi$ after twirling with the RF state $\rho_\text{RF}$ is
\be
t= \frac{1}{2}\tr \left|\cG\left[(\proj{\Psi}-\Psi)\otimes \rho_\text{RF} \right]\right|.
\label{eq:trace_distance_macro_states}
\ee
The state of the RF can always be decomposed into a convex sum of pure RF states $\ket{\text{RF}_i}=\sum_n r_n^{(i)} \ket{n}$, i.e., 
$\rho_\text{RF} = \sum_i q_i \proj{\text{RF}_i}$.  As the trace distance is jointly convex, it follows that $t \leq \sum_i q_i t^{(i)}$ where
\be
t^{(i)} \equiv  \frac{1}{2}\tr \left|\cG\left[(\proj{\Psi}-\Psi)\otimes \proj{\text{RF}_i} \right]\right|.
\label{eq:convex_trace}
\ee

After twirling the state $\ket{\Psi}\ket{\text{RF}_i}$ decomposes into block-diagonal form
\be
\cG\left[\proj{\Psi}\otimes\proj{\text{RF}_i}\right]=\bigoplus_k P_{k,i} \proj{\Psi_{k,i}},
\label{eq:twirled_macro_state}
\ee
where 
\begin{align}\nonumber
\ket{\Psi_{k,i}}&=\frac{\psi_{k-n} r_n^{(i)}}{\sqrt{P_{k,i}}} \ket{k-n}_\text{S}\ket{n}_\text{RF}\\
&\equiv \sum_n \lambda^{(k,i)}_n\ket{n^{(k)}}
\label{eq:psi_i}
\end{align}
and $P_{k,i} \equiv \sum_n (\psi_{k-n} r_n^{(i)})^2$.  The corresponding semi-classical state, after twirling with the RF, reads  
\begin{equation}
\cG\left(\Psi\!\otimes\! \proj{\text{RF}_i}\right)=\bigoplus_k P_{k,i} \rho_{k,i}
\label{eq:twirled_semi-classical}
\end{equation}
where
\be
\rho_{k,i}=\sum_{n,m} \lambda^{(k,i)}_n \lambda^{(k,i)}_m e^{-\frac{(n-m)^2}{N^{2\beta}}}\ketbra{n^{(k)}}{m^{(k)}}.
\label{eq:rho_i}
\ee
Using the bound $t\leq \sqrt{1-F^2}$ between the trace distance and fidelity~\cite{MikeIke} one obtains
\be
t^{(i)}\leq \sum_k P_{k,i}\sqrt{1 - \bra{\Psi_{k,i}}\rho_{k,i}\ket{\Psi_{k,i}}}.
\label{eq:tracedist_i}
\ee
Using the inequality $e^{-d^2}\geq 1-d^2$ and substituting Eqs.~(\ref{eq:psi_i},~\ref{eq:rho_i}) into Eq.~\eqref{eq:tracedist_i} one obtains
\be
t^{(i)}\leq \sum_k P_{k,i}\sqrt{\sum_{n,m} (\lambda^{(k,i)}_n)^2 (\lambda^{(k,i)}_m)^2 \frac{(n-m)^2}{2N^{2\beta}}}.
\label{eq:tracedist_i_2}
\ee
Expanding the quadratic term inside the square root of Eq.~\eqref{eq:tracedist_i_2}, we may express the entire sum as the variance of the distribution 
$(\lambda^{(k,i)}_n)^2$, yielding 
\be
t^{(i)}\leq \sum_k P_{k,i}\sqrt{\frac{\text{Var}\Big((\lambda^{(k,i)}_n)^2\Big)}{N^{2\beta}}}.
\ee
Finally, repeated use of Jensen's inequality and the normalization condition $\sum_n \psi_{k-n}^2 =1$ (see App.~\ref{app:macro_states}) gives the desired result
\be
t\leq \sqrt{\frac{\sum_i  q_i \text{Var}\Big( (r_n^{(i)})^2\Big)}{N^{2\beta}}}.
\label{eq:final_trace_dist}
\ee
Thus the trace distance vanishes whenever the average variance of the RF state $\rho_\text{RF}$ scales slower than $N^{2\beta}$. For the classical states of the RF considered above, and for any state after the action of generic local noise~\cite{IP}, the average variance scales linearly with the size of the RF, $\sum_i  q_i \text{Var}\Big( (r_n^{(i)})^2\Big)\approx M$. Hence, we have shown that in order to certify the presence of quantum coherence on the scale $|n-m|>N^\beta$ one requires a classical RF of size $M> N^{2\beta}$. In particular, genuine macroscopic coherences $|n-m|=O(N)$ require a \emph{quadratically} larger RF. 

We stress that our result holds only for RF states whose quantum features can be explained as the accumulation of $\mathcal{O}(N)$ macroscopic quantum effects.  If the measurement device is allowed to possess macroscopic quantum features as well, then the requirements on its size change.  For example, in the photonic case, it can be shown that an RF prepared in the so called sine state~\cite{Berry00}
\be
\ket{\text{RF}}=\sqrt{\frac{2}{N+2}}\sum_{n=0}^N\sin\left(\frac{(n+1)\pi}{N+2}\right)
\label{eq:sine_state}
\ee
only needs to scale linearly with the size of the macroscopic superposition in order to distinguish the latter form the corresponding mixture in a single shot.

\section{Discussion}
\label{sec:Discussion}

We have shown how in order to distinguish a macroscopic superposition from an incoherent mixture in a single-shot, using classical measuring devices, the size of such devices needs to be quadratically larger than the size of the macroscopic superposition in question.  In what follows we relax our requirement for single-shot certification (Sec.~\ref{sec:probability}), and classical measuring devices (Sec.~\ref{sec:macrorfs}) and determine how relaxing these conditions affects the successful certification of macroscopic superpositions, and/or the size of the corresponding measuring device. In Sec.~\ref{sec:CATS} we discuss the ramifications of our findings for certifying macroscopic quantum states.

We note that a similar result was obtained in a different context by Kofler and Brukner~\cite{Kofler10}.  In particular, Kofler and Brukner show that in order to observe violations of local realism for bigger and bigger spins, one requires measuring devices with larger and larger accuracy (or finer and finer coarse-graining). The authors then connect the resolution of measurement devices with their size by applying Heisenberg's uncertainty relations at the macroscopic scale and using relativistic causality. Here, we adopted a more direct approach by internalizing the states of classical measurement devices in the quantum formalism and invoking fundamental symmetries.
\subsection{Repeating the measurement}
\label{sec:probability}

Our analysis so far was centered on the assumption that we are given only a single copy of the macroscopic state and thus had to ensure that we are able to distinguish it from the incoherent mixture, with a high probability of success, in a single experimental run.  All cases we considered exhibited the same behavior for the probability of success, namely  that for a classical RF of size $M$ the trace distance between a superposition state of size $N$ and the corresponding mixed state is generically given by
\be
t=\frac{1}{2}\exp\left(-\frac{N^2}{8 M}\right).
\ee 
If we want the probability of success to be high, then $M>N^2$.  

However, in practice one is of course not bound to a single shot measurement and one can overcome the limitation set by the the size of the RF by repeating the experiment sufficiently many times. A legitimate question then is how many repetitions are required in order to confirm that a superposition was prepared given our RF is of size $M<N^2$. The answer to this question is provided by the Chernoff bound of Eq.~\eqref{eq:Chernoff}. 

Consider a scenario where one performs a series of experiments with superposition states of larger and larger $N$, and with a   corresponding RF whose size is given by $M=c N^{2-\varepsilon}$, i.e., the RF also increases in subsequent experiments as  as function of $N$.  We ask how the number of repetitions $n$, necessary to witness the superposition from the mixture, grows as a function of the size of the macroscopic quantum state $N$. First, note that if the size of the RF can grow at least quadratically with $N$, i.e., $\varepsilon \leq0$, then the trace distance in fact decreases with $N$, so that large superpositions can be certified without the need to increase the number of repetitions. But for practical reasons we are more interested in the case $\varepsilon >0$ (e.g. $\varepsilon =2$ if the size of the RF is limited), where the resources are limited such that a quadratic growth of the RF is not possible. In this case the number of repetitions, $n$, has to increase in order to compensate for a decreasing trace distance, and one can easily see from Eq.~\eqref{eq:Chernoff} that for a given constant probability of error, $\lim_{n\to \infty}P_\text{Err}^{(n)} \leq p$, one requires that
\be
n \geq  2\log\left(\frac{1}{p}\right) \exp\left(\frac{N^\varepsilon}{8 c}\right).
\ee
In other words, the required number of repetitions blows up exponentially with $N$, so the experiment becomes infeasible quite fast.
\subsection{Implications and examples}
\label{sec:CATS}

We have demonstrated that when describing macroscopic superposition states of several types---having larger and larger size---one eventually has to abandon the 
abstraction of an idealized, classical, RF. In particular, we have shown how the task of distinguishing the superposition of a certain size from the corresponding 
mixture leads to a requirement on the size of the classical RF. We now quantitatively illustrate this relation with two examples. 

First, consider a simple model for the cat in Schr\"odinger's  thought experiment consisting of $N$ spins in a superposition state of Eq.~\eqref{eq:spinmacro}. For simplicity  let us assume that each spin is carried by a water molecule.  For a cat of mass $\approx$ 3 kg, and with a single water molecule having mass $m_0= 18\, N_A = 3\cdot 10^{-26} \text{kg}$, our cat is comprised of roughly $N= 10^{26}$ water molecules.  Our results imply that in order to certify such a cat in the superposition state of Eq.~\eqref{eq:spinmacro} in a single shot, would require a RF of size $M\approx N^2 = 10^{52}$.  Assuming that the RF is also made entirely of water molecules  implies 
\be
m_\text{RF}\geq N^2 m_0 \approx 3 \cdot 10^{25}\, \text{kg}\approx 50 M_\oplus,
\ee 
where $M_\oplus$ is the mass of the Earth. This clearly indicates the practical impossibility to perform such an experiment.

Second, consider the case of delocalized mass and let us again assume that planet Earth plays the role of the RF. For simplicity we use a model similar to the one above where 
the RF is a crystal of $K$ water molecules, each of which is localized in a Gaussian wave-packet of width $\sigma_0 \approx 10 \text{\r{A}} = 10^{-9} \text{m}$--- roughly the size of an ice crystal cell. Requiring the exponent in Eq.~\eqref{eq:massive size} to be $\leq 1$ shows that even an RF of such astronomical size only allows to observe a massive superposition state of Eq.~\eqref{eq:positionmacro} for 
\be
L \,m \leq 10^{-9}\, \text{kg m}.
\ee

This model leads us to conclude that in order to observe a single water molecule delocalized over one meter $L =1 \,\text{m}$ the requisite ``water crystal'' RF  must have a mass $M = K m_0 \approx 3 \, \mu\text{g}$ which is not a severe practical limitation. But
for an object of mass $m=1 \,\text{g}$ and the RF of the size of the Earth $K=M_\oplus/ m_0 \approx 2\cdot~10^{50}$ this gives a very harsh restriction $L \leq 1  \,\mu\text{m}$. 
\section{Many copies and relative degrees of freedom}
\label{sec:macrorfs}

We have seen how symmetry arguments lead to severe restrictions on the size of the measurement device that allow to observe macroscopic superpositions of the type of Eqs.~(\ref{eq:photonmacro},~\ref{eq:spinmacro},~\ref{eq:positionmacro}). Here, we will show how such limitations can be circumvented in two ways: by manipulating two copies of the superposition states, or by preparing superpositions in relative degrees of freedom. 

\subsection{Two copies of macroscopic superposition}

Let us now suppose that one is able to prepare and coherently manipulate two copies of the macroscopic superposition states simultaneously. In this case the macroscopic superpositions play the role of a RF for each other, and it becomes possible to distinguish two copies of a superposition from two mixtures without the need of an additional RF as we now show.

For the purposes of illustration let us first consider the photonic superposition states of Eq.~\eqref{eq:photonmacro}. The task now is to distinguish the superpositions $\ket{\Psi}\ket{\Psi}$ from two mixtures $\Psi\otimes\Psi$ without a external RF state. Recalling that the restrictions imposed by time-translation symmetry are equivalent to replacing the global states of a closed system by their $G$-twirled version 
\begin{align}\nonumber
\cG[\proj{\Psi}\otimes\proj{\Psi}]&=\frac{\proj{00}}{4}+\frac{\proj{NN}}{4}+\frac{\sigma_N}{2} \\
\cG[\Psi\otimes\Psi]&=\frac{\proj{00}}{4}+\frac{\proj{NN}}{4}+\frac{\rho_N}{2},
\label{eq:macrorf}
\end{align} 
where
\begin{align}\nonumber
\sigma_N&=\left(\begin{matrix}
1&1\\1&1\end{matrix}\right)\\
\rho_N&=\left(\begin{matrix}
1&0\\0&1\end{matrix}\right),
\end{align}
are expressed in the basis $\{\ket{0}\ket{N},\ket{N}\ket{0}\}$.  It is trivial to show that the trace distance between the states of Eqs.~\eqref{eq:macrorf} is $t=1/4$. This means that the probability to correctly distinguish the two states in a single shot is as high as $3/4$, and it quickly converges to one when the experiment is repeated. The situation is similar for the case of spins and for massive superpositions, although for the case of spins care must be taken as one has to consider the entire $G$-twirling map over all possible orientations (see App.~\ref{app:orientationtwirling}).

\subsection{Macroscopic superpositions in relative degrees of freedom}

These observations are consequences of the fact that superposition in  relative (or interal) degrees of freedom are not affected by global symmetries. Hence, for example a photonic two-mode superposition of the form$\frac{1}{\sqrt{2}}\left(\ket{N,0} + \ket{0,N}\right)$
might suffer from various technical limitations, but does not require an external phase RF in order to be prepared. 

A similar argument holds for the case of position with two objects of masses  $m_1$ and $m_2$; the phase space of the relative degree of freedom $x = x_1 - x_2$ and $p =\frac{m_1 p_1 - m_2 p_2}{m_1 +m_2}$ is not affected by the twirling map for the center of mass degree of freedom. These two objects can be, for example, a heavy molecule and the double slit in an interference experiment. However if the mass of the molecule is big enough, this indicates that the state of the double slit has to be involved in the final measurement.

In the case of spins one can also construct a macroscopic superposition of two states both with $J=0$ and  thus invariant under twirling with respect to the orientation group. To do so consider the $J=0$ subspace for four spin-$\frac{1}{2}$ particles. This subspace is spanned by two states $\ket{\heartsuit} =\ket{\Psi^-_{12}}\otimes \ket{\Psi^-_{34}}$ and $\ket{\diamondsuit}=\frac{1}{\sqrt{2}}\left(\ket{\Psi^-_{13}}\otimes \ket{\Psi^-_{24}}+\ket{\Psi^-_{14}}\otimes \ket{\Psi^-_{23}}\right)$, where the $ \ket{\Psi^-_{ij}}$ is the singlet states carried by qubits $i$ and $j$. These states are orthogonal and are both invariant under global rotations of al the qubits. Hence, the $4 N$ particle  superposition state $\frac{1}{\sqrt{2}}\left(\ket{\heartsuit}^{\otimes N} +\ket{\diamondsuit}^{\otimes N} \right)$ is invariant under rotations and does no require an external RF for spatial orientation. As before, this can be understood as a superposition in an internal (relative) degree of freedom, as the two states $\ket{\heartsuit}$ and $\ket{\diamondsuit}$ are related by the action of the permutation group.

\section{Conclusion}
\label{sec:Conclusion}
We have shown that in order to distinguish macroscopic superpositions from their incoherent mixtures in a single-shot measurement one requires classical measuring devices that are quadratically larger.  Specifically, we have highlighted how relevant degrees of freedom of classical measuring devices, such as size, orientation, position, velocity etc, play a pivotal role acting as appropriate RFs.  By incorporating the latter within the quantum formalism we are able to provide a quantification of the size of macroscopic quantum states in terms of the size of the corresponding classical measuring device needed to distinguish it.  
In addition, we have shown how the requirement of quadratically larger size for classical devices can be relaxed if we are allowed to repeat the experiment. However, the necessary number of repetition then increases exponentially with the size of the superposition state, making this route practically unfeasible.  By using one macroscopic state as a RF for a second copy of the state, we can significantly reduce the number of repetitions needed in order to distinguish the states from their classical mixtures.

Whereas we have focused solely on certifying macroscopic superpositions, we believe that the arguments presented here should hold for the preparation of such states. A more direct connection of our quadratically larger RF for the preparation of such states as well is the subject of future work, along with an extension to other symmetry groups, such as the Lorentz and Poincare groups of special and general relativity, as well as for more specific situations where both the preparation and certification of macroscopic states are performed by the same device.

Finally, we also showed how the limitations are circumvented by macroscopic superposition in relative (or internal) degrees of freedom, that are invariant under global symmetries and thus do not require an external RF to be defined. 

\acknowledgements
This work was supported by  Spanish MINECO FIS2013-40627-P, IJCI-2015-24643, and Generalitat de Catalunya CIRIT 2014 SGR 966.GR 966. (MS), the Austrian Science Fund (FWF):  P28000- N27, SFB F40-FoQus F4012-N16, and the Swiss National Science Foundation grant P300P2\_167749.
\appendix 
\section{Quantum mechanical treatment of reference frames}
\label{app:G-twirling}

In this appendix we briefly review the connection between symmetry with respect to a group of transformations and the lack of a RF.  We then show how to incorporate RFs within the quantum mechanical formalism. We will only focus on the relevant ingredients needed for this work; the detailed treatment of the problem  can be found in~\cite{Bartlett07}. 

Let $\cH^{(\mathrm{S})}$ denote the state space of a quantum mechanical system.  The action of a symmetry group $G$ on $\cH^{(\mathrm{S})}$ is represented by a set of transformations $\{U(g)\,:\, g\in G\}$.  We write $U:G\to\cH^{(\mathrm{S})}$, and for ease of exposition we shall focus our discussion to finite dimensional Hilbert spaces, and to unitary representations of compact Lie groups.  Now in the presence of a symmetry associated with the group $G$ the only allowable operations, $A$, are those that satisfy $[A, U(g)]=0,\,\forall\, g\in G$.  We will now show that this restriction is formally equivalent to the lack of a RF associated with $G$ relative to which the states of a physical system are defined.

Specifically consider performing the allowable von Neumann measurement $\proj{k}$. As $\proj{k}$ is an allowable operation, it follows that $\proj{k}=U(g)^\dagger\proj{k} U(g)$.  Moreover, as this is true for all $g\in G$, it follows that 
\begin{align}\nonumber
\proj{k}&=\int_{g\in G}\,\mathrm{d}g\, U(g)^\dag\,\proj{k}\, U(g)\\
&\equiv\cG[\proj{k}],
\label{app:G-map}
\end{align}  
where $\mathrm{d}g$ denotes the uniform Haar measure of the group $G$.  Now consider the corresponding probability of obtaining the outcome associated to $\proj{k}$ given an initial state $\ket{\psi}$, not necessarily obeying the symmetry associated to $G$.  One easily obtains
\begin{align}\nonumber
p_k&=\mathrm{Tr}\left[\proj{k}\, \proj{\psi}\right]\\ \nonumber
&=\mathrm{Tr}\left[\cG[\proj{k}] \, \proj{\psi}\right]\\
&=\mathrm{Tr}\left[\proj{k}\, \cG[\proj{\psi}]\right],
\label{app:symmetry}
\end{align}
where we have used the fact that $\cG^\dag=\cG$.  Eq.~\eqref{app:symmetry} shows that the restrictions imposed by symmetry on the allowable operations are operationally indistinguishable from the situation where the state of the physical system is described by $\cG[\proj{\psi}]$ as opposed to $\ket{\psi}$.  

To see the connection with RFs, consider the following situation.  Alice and Bob each hold their own respective RFs associated with the group of transformations $G$. However, their RFs are not aligned, but rather related by some $g\in G$.  If Alice prepares the state $\ket{\psi}$ relative to her RF, then Bob's description of the state is 
$U^\dag(g)\ket{\psi}$, to account for the relation between his and Alice's RFs.  Now suppose that Alice and Bob are completely ignorant as to which $g\in G$ relates their corresponding RFs.  This is equivalent to assuming that Bob does not possess the requisite RF at all.  In this case  Bob's description of the system is given by the state
\begin{align}\nonumber
\rho&=\int_{g\in G}\, \mathrm{d}g\, U(g)^\dag\,\proj{\psi}\, U(g)\\
&=\cG[\proj{\psi}].
\label{app:lackofrf}
\end{align}  
Hence, the presence of symmetry associated with a group of transformations $G$ is operationally equivalent to the lack of a RF associated with $G$. Indeed, the restrictions can be partially lifted if Alice and Bob perform RF alignment~\cite{Bagan04, Chiribella04a, Chiribella04b, Chiribella05}, where one of the parties encodes their local RF in the state of a physical system---known as a quantum \defn{token} of a RF---which the other party measures in order to extract as much information about the RF as possible.  Clearly a good token of a RF is one whose orbit under the action of the group $G$ is non-trivial.  Denoting by $\Gamma:G\to\cH^{(\mathrm{RF})}$ the unitary representation of $G$ onto the state space of the RF token, it follows that the  more distinguishable the set of states  $\{\ket{\mathrm{RF}(g)},\,\forall g\in G\}$ are from each other the more asymmetric the RF token state $\ket{\text{RF}}\in\cH^{(\mathrm{RF})}$ is with respect to the symmetry group $G$ and thus the better its ability to act as a RF. 

The above discussion highlights the crucial role played by RFs in the description of physical systems.  Indeed, whenever one writes down a pure state of a quantum system one \defn{implicitly} assumes a RF; saying that a system is in the $+1$ eigenstate of $\sigma_z$ assumes the notion of a $z$-direction. More often than not, the RF is assumed to be an idealized system external to, and independent from, the theory used to describe the systems in question.  If, however, we adopt the point of view that RFs are physical systems themselves then we need to treat them on an equal footing as all other systems.       

To that end, let us now describe the situation where we incorporate the requisite RF within our quantum mechanical description.  That is consider the 
state $\ket{\text{RF}}\otimes\ket{\psi}\in\cH^{(\mathrm{R})}\otimes\cH^{(\mathrm{S})}$.  Now consider a hypothetical RF relative to which the joint state $\ket{\text{RF}}\otimes\ket{\psi}$ is defined.  Changing this hypothetical RF by $g\in G$ is equivalent to performing the unitary transformation $\Gamma(g)^\dagger\otimes U(g)^\dagger(\ket{\text{RF}}\otimes\ket{\psi})$. However, as such a global RF is not physically available, the joint state of RF-plus-system is 
\begin{align}\nonumber
\cG[\proj{\text{RF}}\otimes\proj{\psi}]&=\int\, \mathrm{d}g \, \left(\Gamma(g)^\dagger\otimes U^\dagger(g)\right)\times\\
&\proj{\text{RF}}\otimes\proj{\psi}\, \left(\Gamma(g)\otimes U(g)\right).
\label{app:doublereptwirling}
\end{align}
Using Schur's lemmas~\cite{Sternberg94} there exists an orthonormal basis relative to which $\Gamma\otimes U$ can be decomposed into several inequivalent and irreducible representations (irreps), $\Lambda^{(\lambda)}$, as follows 
\begin{equation}
\Gamma(g)\otimes U(g)\equiv\bigoplus_\lambda \Lambda^{(\lambda)}(g)\otimes I^{(\lambda)},
\label{app:irreps}
\end{equation}
where the identity operator $I^{(\lambda)}$ has dimension equal to the number of times the irrep $U^{(\lambda)}$ appears in the decomposition of $\Gamma\otimes U$. Consequently, the Hilbert space $\cH^{(\mathrm{RF})}\otimes\cH^{(\mathrm{S})}$ can be written in the same basis as
\begin{equation}
\cH^{(\mathrm{R})}\otimes\cH^{(\mathrm{S})}\equiv\bigoplus_\lambda\cH^{(\mathrm{R}+\mathrm{S})}_\lambda\equiv \bigoplus_\lambda \cM^{(\mathrm{R}+\mathrm{S})}_\lambda\otimes \cN^{(\mathrm{R}+\mathrm{S})}_\lambda,
\label{app:reducibility}
\end{equation}
where $\cH^{(\mathrm{R}+\mathrm{S})}_\lambda$ are the invariant subspaces of $\cH^{(\mathrm{R})}\otimes\cH^{(\mathrm{S})}$ under the action of $\Gamma\otimes U$; for any $\ket{\psi}\in\cH^{(\mathrm{R}+\mathrm{S})}_\lambda$, $\Gamma(g)\otimes U(g)\ket{\psi}\in\cH^{(\mathrm{R}+\mathrm{S})}_\lambda,\,\forall\, g\in G$.  The state spaces $\cM^{(\mathrm{R}+\mathrm{S})}_\lambda\,(\cN^{(\mathrm{R}+\mathrm{S})}_\lambda)$ are the spaces upon which the  irreps $\Lambda^{(\lambda)}$ act non-trivially (trivially) and are known as the \defn{carrier} (\defn{multiplicity}) spaces respectively.  For example, in the case where $G=u(1)$---the symmetry group of time-translations that corresponds to a RF for phase $\theta\in(0,2\pi]$---the carrier spaces are the one dimensional spaces of total photon number $\hat{N}=\hat{N}_{\mathrm{RF}}+\hat{N}_{\mathrm{S}}$, $\cM_n^{(\mathrm{RF}+\mathrm{S})}\equiv\mathrm{span}\{\ket{n}\equiv\ket{n}_{\mathrm{RF}}\otimes\ket{n}_{\mathrm{S}}\}$, and the multiplicity spaces are simply all possible ways of partitioning the total photon number into two parts.  In the case where $G=su(2)$---the symmetry group of rotations that corresponds to an orientation RF---the carrier spaces are given by $\cM_J^{(\mathrm{RF}+\mathrm{S})}=\mathrm{span}\{\ket{J,M}\}$ where $\ket{J,M}$ is the joint eigenstate of the total angular momentum operator $J^2=J_x^2+J_y^2+J_z^2$, and $J_z$, and the multiplicity spaces are the various different ways two quantum systems of angular momentum $j_{\mathrm{RF}}$ and $j_{\mathrm{S}}$ can couple to give rise to a total angular momentum $J$.    

In the basis of Eq.~\eqref{app:reducibility} the $G$-twirling map of Eq.~\eqref{app:doublereptwirling} reads~\cite{Bartlett07}
\begin{align}\nonumber
\cG[\proj{\text{RF}}\otimes\proj{\psi}]&=\bigoplus_\lambda \left(\cD_{\cM^{(\mathrm{R}+\mathrm{S})}_\lambda}\otimes\cI_{\cN^{(\mathrm{R}+\mathrm{S})}_\lambda}\right)\circ\\
&\cP_\lambda[\proj{\text{RF}}\otimes\proj{\psi}],
\label{app:Gtwirledreducible}
\end{align} 
where $\cP_\lambda[\rho]\equiv \Pi_\lambda\rho\Pi_\lambda,\, \rho\in\cB(\cH^{(\mathrm{R})}\otimes\cH^{(\mathrm{S})})$ with $\Pi_\lambda$ the projector onto $\cH_\lambda^{(\mathrm{R}+\mathrm{S})}$, $\cD_{\cM^{(\mathrm{R}+\mathrm{S})}_\lambda}[A]=\frac{\mathrm{Tr}(A)}{|\cM^{(\mathrm{R}+\mathrm{S})}_\lambda |}\eins_{\cM^{(\mathrm{R}+\mathrm{S})}_\lambda},\, \forall \, A\in\cB(\cM^{(\mathrm{R}+\mathrm{S})}_\lambda)$ with $|\cM^{(\mathrm{R}+\mathrm{S})}_\lambda |$ the dimension of the carrier space $\cM^{(\mathrm{R}+\mathrm{S})}_\lambda$, and  $\cI_{\cN^{(\mathrm{R}+\mathrm{S})}_\lambda}[A]=A,\, \forall\, A\in\cB(\cN^{(\mathrm{R}+\mathrm{S})}_\lambda)$.  

The intuition behind the particular decomposition of the total space $\cH^{(\mathrm{R})}\otimes \cH^{(\mathrm{S})}$ is as follows.  The carrier space $\cM^{(\mathrm{R}+\mathrm{S})}_\lambda$ corresponds to the global degrees of freedom of the joint RF-plus-system, whereas the multiplicity space $\cN^{(\mathrm{R}+\mathrm{S})}_\lambda$ corresponds to the relative  degrees of freedom.  It is in the latter that all relevant properties of interest regarding the relationship of the system to the RF preside.

\section{Measurements under conservation laws}
\label{app:measurement}

In this appendix we provide the detailed computations of how well one can perform a measurement in the eigenbasis of $J_x$ on a spin $N/2$ system under energy conservation imposing that only projections onto the eigenbasis of $J_z$ are allowed.  The measurement can be achieved by utilizing a bright laser pulse in the coherent state $\ket{\alpha}$, and the Jaynes-Cummings interaction   
\begin{equation}
U=e^{\gamma(J_+a-J_-a^\dagger)},
\label{app:JCN}
\end{equation}
to perform a rotation by $\pi/2$ about the $y$-axis followed by a measurement of $J_z$.  In Eq.~\eqref{app:JCN}, $a$ and $a^\dagger$ are the annihilation and creation operators, $\gamma$ is the interaction strength, and $J_\pm$ are the corresponding raising and lowering operators of a spin-$N/2$ system.  

For ease of exposition, we will make use of the Schwinger representation
\begin{align}\nonumber
J_x&=\frac{1}{2}\left(b^\dagger c+bc^\dagger\right)\\ \nonumber
J_y&=\frac{1}{2i}\left(b^\dagger c-bc^\dagger\right)\\ 
J_z&=\frac{1}{2}\left(b^\dagger b-c^\dagger c\right),
\label{app:Schwinger}
\end{align}
of spin operators in terms of the operators of two bosonic modes.  The corresponding eigenstates of the $J_z$ can be written in terms of the states of two bosonic modes in the following way 
\begin{equation}
\ket{N/2,m}=\ket{m}\otimes\ket{N-m}
\label{app:bosonicstates}
\end{equation}
so that $J_z\ket{N/2,m}=\frac{2m-N}{2}\ket{N/2,m}$.  

Writing $\ket{\alpha}=D(\alpha)\ket{0}$, where $D(\alpha)$ is the displacement operator, and noting that 
\begin{equation}
e^{\gamma(J_+a-J_-a^\dagger)}D(\alpha)=D(\alpha)e^{i2\alpha \gamma J_y}e^{\gamma(J_+a-J_-a^\dagger)},
\label{app:displacement}
\end{equation}
we see that upon tracing out the optical mode results in the measurement
\begin{align}\nonumber
\cE[E_m]&=\mathrm{tr}_{RF}\left[U(\proj{\alpha}\otimes E_m)U^\dagger\right]\\ \nonumber
&=\mathrm{tr}_{RF}\left[D(\alpha)e^{\gamma(i2\alpha J_y+(J_+a-J_-a^\dagger))}(\proj{0}\otimes E_m)\right.\\ \nonumber
&\left. \times e^{-\gamma(i2\alpha J_y+(J_+a-J_-a^\dagger))}D(-\alpha)\right]\\ \nonumber
&=\mathrm{tr}_{RF}\left[e^{\gamma(i2\alpha J_y+(J_+a-J_-a^\dagger))}(\proj{0}\otimes E_m)\right.\\
&\left. \times e^{-\gamma(i2\alpha J_y+(J_+a-J_-a^\dagger))}\right],
\label{app:noisymeasurement}
\end{align}
where $E_m=\proj{N/2,m}$. Choosing $\gamma=\frac{\pi}{4\alpha}$ results in the desired rotation about the $y$-axis up to a small perturbation that leads to noise described by the CPTP map with Kraus operators   
\begin{align}
K_n=\bra{n}e^{(i\frac{\pi}{2}J_y+\gamma(J_+a-J_-a^\dagger))}\ket{0},
\label{app:krauss1}
\end{align}
and vanishes in the limit $\gamma N\to 0$.  

In order to quantify the distance of the resulting measurement from the ideal one we Taylor expand Eq.~\eqref{app:krauss1} around $\gamma=0$ yielding
\begin{equation}
K_n=\bra{n}\left(\eins+\gamma O_1+\gamma^2O_2\right)\ket{0}+\mathcal{O}(\gamma^3 O_3),
\label{app:krauss2}
\end{equation}
where 
\begin{align}\nonumber
O_1&=\int_0^1\mathrm{d}x\,e^{ix\frac{\pi}{2}J_y}\left(J_+a-J_-a^\dagger\right)e^{-ix\frac{\pi}{2}J_y}\\ \nonumber
O_2&=\int_0^1\mathrm{d}x\int_0^x\mathrm{d}y\,e^{iy\frac{\pi}{2}J_y}\left(J_+a-J_-a^\dagger\right)e^{i(x-y)\frac{\pi}{2}J_y}\\ 
&\times \left(J_+a-J_-a^\dagger\right)e^{-ix\frac{\pi}{2}J_y}.
\end{align}
We note that, since $\lvert\lvert[J_+a-J_-a^\dagger]\rvert\rvert\leq N$, $\|\gamma^3O_3\|\leq \gamma^3 N^3$ and thus we only consider terms up to second order in $\gamma$.  As $J_\pm=J_x\pm iJ_y$, using the property
\begin{equation}
e^{i\theta J_y}J_\alpha e^{-i\theta J_y}=\sum_{\beta\in(x,y,z)}R_y(\theta)_{\alpha\beta}J_\beta,
\label{app:rotations}
\end{equation}
and noting that, up to second order in $\gamma$, only the Kraus operators $K_0$ and $K_1$ are of relevance, one obtains
\begin{align}\nonumber
K_0&=\eins-\frac{2\gamma^2}{\pi^2}\mathbf{L}^\dagger \, A\, \mathbf{L}\\
K_1&=\frac{-2\gamma}{\pi} \mathbf{L}^\dagger\cdot \mathbf{v},
\label{app:krauss3}
\end{align}
where $\mathbf{L}=(J_z,\, J_+,\, J_-)^T$ and 
\begin{align}\nonumber
A&=\left(\begin{matrix}
1 & 2-\frac{3\pi}{4} & \frac{\pi}{4}\\
\frac{\pi-4}{4}& \frac{(\pi-2)^2}{16} &\frac{20-\pi(\pi+4))}{16}\\
\frac{\pi+4}{4}& \frac{-12-\pi(\pi-4)}{16} & \frac{(\pi+2)^2}{16}
\end{matrix}\right),\\
\mathbf{v}&=\left(\begin{matrix}
1\\
\frac{2-\pi}{4}\\
\frac{2+\pi}{2}
\end{matrix}\right).
\label{app:krauss4}
\end{align}

We quantify the distance between the ideal measurement and the noisy one given in Eq.~\eqref{app:noisymeasurement}, by calculating the average overlap between the ideal projectors $\proj{N/2,m}$ and the corresponding noisy ones $\cE[\proj{N/2,m}]$.  The latter is given by $\bar{f}\equiv\frac{1}{N+1}\sum_{n=0}^{N}\mathrm{tr}[\proj{N/2,m},\cE[\proj{N/2,m}]]$ which reads
\begin{align}\nonumber
\bar{f}&=\frac{1}{N+1}\sum_{m=0}^{N}\left\lvert\bra{\frac{N}{2},m}K_0\ket{\frac{N}{2},m}\right\rvert^2\\
&+\left\lvert\bra{\frac{N}{2},m}K_1\ket{\frac{N}{2},m}\right\rvert^2.
\label{app:overlap1}
\end{align}
Observing that only terms involving $J_z^2$, $J_+J_-$ and $J_-J_+$ contribute in the first summand of Eq.~\eqref{app:overlap1} and defining 
\begin{align}\nonumber
z_m&=\frac{2m-N}{2}\\ \nonumber
r^+_m&=\sqrt{(m+1)(N-m)}\\
r^-_m&=\sqrt{m(N-(m-1))},
\label{app:defs}
\end{align}
Eq.~\eqref{app:overlap1} reads
\begin{align}\nonumber
\bar{f}&=1-\frac{4\gamma^2}{\pi^2(N+1)}\left(\sum_{m=0}^N (A_{11}-v_1^2)z_m^2 \right.\\ \nonumber
&\left.+A_{22}r^-_{(m+1)}r^+_m+A_{33}r^+_{(m-1)}r^-_m\right)\\
&=1-\frac{\gamma^2N(N+2)(4+\pi^2)}{12\pi^2}+\mathcal{O}(\gamma^3N^3)
\label{app:overlap2}
\end{align}
where $A_{ij}$ ($v_i$) denotes the elements of matrix $A$ (vector $\mathbf{v}$) in Eq.~\eqref{app:krauss4}. 
Recalling that $\gamma=\pi/4\alpha$ gives 
\begin{equation}
\bar{f}=1-0.07\left(\frac{N(N+2)}{\alpha^2}\right)+\mathcal{O}(\gamma^3N^3).
\label{app:finalresult}
\end{equation}
Thus, the error we incur in performing the ideal measurement along the $x$ direction for a spin $N/2$ system scales inversely proportional with the average number of photons present in our coherent laser beam.  For the case of a spin-$1/2$ system, the average error reads $\bar{f}=1-0.22\alpha^{-2}$. 

It is instructive to consider precisely the effect of Eq.~\eqref{app:noisymeasurement} on the projectors, $\proj{N/2,m}$.  The resulting measurement is given by 
\begin{align}\nonumber
&\cE[\proj{N/2,m}]=\proj{N/2,m}\\ \nonumber
&+\frac{1}{4\alpha^2}\mathbf{v}^\mathrm{T}\mathbf{L}\proj{N/2,m}\mathbf{L}^\dagger\mathbf{v}\\\nonumber
&-\frac{1}{16\alpha^2}\left(\mathbf{L}^\dagger A\mathbf{L}\proj{N/2,m}+\proj{N/2,m}\mathbf{L}^\dagger A\mathbf{L}\right)\\
&=\proj{N/2,m}+\Delta_m+\Delta_m^\mathrm{T},
\end{align}
where the matrix $\Delta_m$ has entries only within the range $(m-2,\ldots,m+2)$ and is explicitly given by 
\begin{widetext}
\begin{equation}
\Delta_m=\frac{-1}{8\alpha^2}\left(
\begin{array}{ccccc}
 0 & 0 & 0 & 0 & 0 \\
 0 & -v_3^2 r^{+2}_{m-1} & 0&0& 0 \\
 A_{23}r^+_{m-2}r^+_{m-1} & \left(A_{13} z_{m-1}+(A_{21}-2v_1v_3)
   z_m\right)r^+_{m-1} & (A_{11}-v_1^2) z_m^2+A_{33}
   r^{+2}_{m-1}+A_{22}r^+2_m & 0& 0 \\
 0 & -2v_2v_3 r^+_{m-1}r^+_m & \left((A_{31}-2 v_1v_2) z_m+A_{12}
   z_{m+1}\right) r^+_m & -v_2^2 r^{+2}_m & 0 \\
 0 & 0 & A_{32} r^+_m r^+_{m+1}& 0 & 0 
\end{array}
\right).
\end{equation}
\end{widetext}
\section{Distinguishing macroscopic superpositions}
\label{app:results}
In this appendix we provide the detailed calculation for discriminating macroscopic superpositions from their incoherent mixtures.  We first consider the photonic case (Appendix~\ref{app:Photonic}), followed by the case of macroscopic superpositions of positional eigenstates (Appendix~\ref{app:Positional}). 

\subsection{Photonic macroscopic superpositions}
\label{app:Photonic}
Our goal is to distinguish the macroscopic superposition of Eq.~\eqref{eq:photonmacro}, from the corresponding incoherent mixture Eq.~\eqref{eq:photonmixture} using the coherent state $\ket{\alpha}$ as a RF.  The joint RF-plus-system is invariant under phase shifts, which in this instance implies that the total photon number $K=N_{\mathrm{RF}}+N_\mathrm{S}$ is conserved.  Hence, superpositions of states of different total photon number are not allowed, as such states require a global RF---a clock---relative to which the phases of such superpositions are defined.

Operationally, this lack of knowledge, is described by a group of transformations $\{V(\theta)\equiv\Gamma^{(\mathrm{RF})}(\theta)\otimes U^{(S)}(\theta);\, \theta\in(0,2\pi]\}$, that describe the relevant phase reference.  Here $\Gamma^{(\mathrm{RF})}(\theta)$ describes the action of the group transformation on the state space, $\cH^{(\mathrm{RF})}$, of the RF, whereas $U^{(S)}(\theta)$ describes the corresponding action on the state space, $\cH^{(\mathrm{S})}$, of the system.  It is important to note that both system and RF are transformed by the same $\theta$.  

The $G$-twirled version of the joint RF-plus-system state (Eq.~\eqref{app:doublereptwirling}) then reads  
\begin{align}\nonumber
\cG[\rho_{\mathrm{RF}}\otimes\sigma_{\mathrm{S}}]&=\frac{1}{2\pi}\int_0^{2\pi}\Gamma^{(\mathrm{RF})}(\theta)\otimes U^{(\mathrm{S})}(\theta)\left[\rho_{\mathrm{RF}}\otimes\sigma_{\mathrm{S}}\right]\\
&\Gamma^{(\mathrm{RF})}(\theta)^\dagger\otimes U^{(\mathrm{S})}(\theta)^\dagger\, \mathrm{d}\theta,
\label{app:u1twirling}
\end{align}
where $\rho_{\mathrm{RF}}\in\cB(\cH^{(\mathrm{RF})})$, $\sigma_{\mathrm{S}}\in\cB(\cH^{(\mathrm{S})})$, and $1/2\pi$ corresponds to the invariant measure of the group and represents our complete lack of knowledge of the global background clock.

Decomposing $\cH^{(\mathrm{RF})}\otimes\cH^{(\mathrm{S})}$ into global and relative photon number states (Eq.~\eqref{app:reducibility}) and using Eq.~\eqref{app:Gtwirledreducible} it follows that
\begin{equation}
\cG[\rho_{\mathrm{RF}}\otimes\sigma_\mathrm{S}]=\bigoplus_{K}p_K\omega_K,
\label{app:blockdiagonal}
\end{equation}
where $p_k=\mathrm{tr}[\rho_{\mathrm{RF}}\otimes\sigma_\mathrm{S} \Pi^{(K)}]$, $\omega_K=\frac{\Pi^{(K)}\rho_{\mathrm{RF}}\otimes\sigma_\mathrm{S} \Pi^{(K)}}{p_K}$ and $\Pi^{K}=\sum_{N_S}\proj{K-N_S}\otimes\proj{N_S}$.

Now consider the case where the RF is given by
\begin{equation}
\ket{\alpha}=\sum_{n=0}^\infty \sqrt{q_n}\ket{n}
\label{app:coherentstate}
\end{equation}
where $q_n=|\alpha|^2n e^{-|\alpha|^2}/n!$.  Under the $G$-twirling map, one can verify that 
\begin{align}\nonumber
&\cG[\proj{R}\hspace{-0.5mm}\otimes\hspace{-0.5mm}\proj{\psi}]=\frac{1}{2}\bigoplus_{K=0}^{N-1}q_K\Pi^{(K)}\bigoplus_{K=N}^\infty \sigma_K\\
&\cG[\proj{R}\hspace{-0.5mm}\otimes\hspace{-0.5mm}\psi]=\frac{1}{2}\bigoplus_{K=0}^{N-1}q_K\Pi^{(K)}\bigoplus_{K=N}^\infty \rho_K,
\label{app:twirledstates}
\end{align}
with 
\begin{align}\nonumber
\sigma_K&=\left(\begin{array}{cc} q_K & \sqrt{q_Kq_{K-N}}\\ \sqrt{q_Kq_{K-N}}& q_{K-N}\end{array}\right)\\
\rho_K&=\left(\begin{array}{cc} q_K & 0\\0& q_{K-N}\end{array}\right).
\label{app:sectors}
\end{align}

We are now ready to compute the trace distance (Eq.~\eqref{eq:tracedistance}) between the two state in Eq.~\eqref{app:twirledstates}.  We note that the trace distance between two states $\rho$ and $\sigma$ can also be expressed as 
\be
t=||\rho-\sigma||=\frac{1}{2}\mathrm{Tr}\left(\sqrt{(\rho-\sigma)^\dagger(\rho-\sigma)}\right).
\label{app:tracedistance}
\ee
As the only non-trivial contributions come from good quantum numbers with $K\geq N$, and by making use of the Gaussian approximation to the Poissonian distribution, the trace distance reads
\begin{align}\nonumber
t&=\frac{1}{\sqrt{8\pi|\alpha|^2}}\int_{x=N}^\infty e^{\frac{-(|\alpha|^2+N-x)^2}{4|\alpha|^2}}e^{\frac{-(|\alpha|^2-x)^2}{4|\alpha|^2}}\mathrm{d}x\\
&=\frac{e^{-\frac{N^2}{8|\alpha|^2}}}{4}\mathrm{Erfc}\left(\frac{N-|\alpha|^2}{\sqrt{2|\alpha|^2}}\right).
\label{tracedistancecoherentstate}
\end{align}
Hence, if $|\alpha|^2>\mathcal{O}(N^2)$, and noting that 
$\mathrm{Erfc}\left(\frac{1-N}{\sqrt{2}}\right)\to2$ for large $N$, the trace distance tends to its maximum value of $1/2$. 

\subsection{Positional macroscopic superpositions}
\label{app:Positional}
We now consider the case of distinguishing the spatial superposition of Eq.~\eqref{eq:positionmacro} from the corresponding  incoherent mixture of Eq.~\eqref{eq:positionmixed} using as a classical measuring device comprised of $K$ localized particles each with mass $m_0$ and the same spread $\sigma_0$ and in the state 
\be
\ket{\mu_i}=\int_{-\infty}^{\infty}\frac{e^{\frac{-(\mu_i-x)^2}{4\sigma_0^2}}}{(2\pi\sigma_0^2)^{\frac{1}{4}}}\ket{x}\mathrm{d}x.
\label{app:localized}
\ee 
Here, $\ket{x}$ are the generalized eigenstates of the position operator $X$, satisfying $\braket{x}{x'}=\delta(x-x')$, $\mu_i$ is the mean position of particle $i$. 

It is convenient to express the state of the device, $\ket{\text{RF}}=\bigotimes_{i=1}^K\ket{\mu_i}$, in terms of the center of mass  position of the $K$ systems, $X=\frac{1}{K}\sum_{i=1}^K (\mu_i-x_i)$, 
\begin{align}\nonumber
&\ket{\text{RF}}=\int_{-\infty}^{\infty}\mathrm{d}X_R\mathrm{d}x_2\ldots\mathrm{d}x_K \left[\frac{1}{(2\pi\prod_{i=1}^K\sigma_0^2)^{\frac{1}{2}}}\right.\\
&\left. e^{\frac{-\left(KX-\sum_{i=2}^K(\mu_i-x_i)\right)^2}{2\sigma_0^2}}\prod_{i=2}^Ke^{\frac{-\left(\mu_i-x_i\right)^2}{2\sigma_0^2}}\right]^{\frac{1}{2}}\ket{X}\otimes\ket{x_2\ldots x_K}.
\label{app:positionrf}
\end{align}
As different states with the same center of mass position are equivalent, defining 
\begin{equation}
\mathrm{RF}({X,\mathbf{x}})=e^{\frac{-\left(KX_R-\sum_{i=2}^K(\mu_i-x_i)\right)^2}{2\sigma_0^2}}\prod_{i=2}^Ke^{\frac{-\left(\mu_i-x_i\right)^2}{2\sigma_0^2}}
\label{app:convolutions}
\end{equation}
and noting that the integrals over $x_2\ldots x_K$ are successive convolutions of Gaussians the state of our classical RF can be written as 
\begin{equation}
\ket{\text{RF}}=\int_{-\infty}^{\infty}\sqrt{\mathrm{RF}(X)}\ket{X}\, \mathrm{d}z,
\label{app:RFstate}
\end{equation}
where
\begin{equation}
\mathrm{RF}(X)=\frac{e^{-\frac{KX^2}{2\sigma_0^2}}}{\sqrt{\frac{2\pi\sigma_0^2}{K}}}
\label{app:positionGaussian}
\end{equation}

Let us focus on the one-dimensional case, where the lack of a momentum RF is described by the group of boosts $v\in(0,\infty)$.  This implies that the center of mass position of the system-plus-device $X_C=\frac{mx+MX}{m+M}$, where the system is of mass $m$ and the RF has a total mass of $M=Km_0$, is conserved.  Observe that unlike the case of a phase reference, the group of co-linear boosts is not compact and thus does not have a uniform Haar measure.  Define $\{\Gamma^{(RF)}(v);\, v\in(0,\infty)\}$ and $\{V^{(S)}(v);\,v\in(0,\infty)\}$ the unbounded unitary representations of the one-dimensional boosts acting on the state spaces, $\cH^{(\mathrm{RF})}$, and $\cH^{(\mathrm{S})}$ of the device and system respectively via   
\begin{align}\nonumber
&\Gamma^{(\mathrm{RF})}(v)\ket{\text{RF}}=\int_{-\infty}^{\infty}\sqrt{\mathrm{RF}(X)}\,e^{\ii vX}\ket{X}\, \mathrm{d}X\,\in\cH^{(\mathrm{RF})}\\
&V^{(\mathrm{S})}(v)\ket{\psi}=\frac{1}{\sqrt{2}}\left(\ket{x-\frac{L}{2}}+e^{\ii vL}\ket{x+\frac{L}{2}}\right)\,\in\cH^{(\mathrm{S})}.
\label{app:unitary}
\end{align}
Then, the $G$-twirled version of the joint system-plus-device state, $\rho_{\mathrm{RF}}\otimes\sigma_{\mathrm{S}}$, that arises due to not knowing $v$ reads
\begin{align}\nonumber
&\cG[\rho_{\mathrm{RF}}\otimes\proj{\Psi}]=\lim_{v\to\infty}\frac{1}{v}\int_0^v e^{\ii v (X+x)}\,\left(\rho_{\mathrm{RF}}\right.\\ \nonumber
&\left.\otimes\proj{\Psi}\right)\, e^{-\ii v(X+x)}\mathrm{d}v\\
&=\bigoplus_{X_C=-\infty}^\infty\left(\begin{matrix} \frac{\mathrm{RF}(X)}{2} &\frac{\sqrt{\mathrm{RF}(X)\mathrm{RF}(X+\frac{mL}{M})}}{2}\\
				     \frac{\sqrt{\mathrm{RF}(X)\mathrm{RF}(X+\frac{mL}{M})}}{2} &  \frac{\mathrm{RF}(X+\frac{mL}{M})}{2} \end{matrix}\right).
\label{app:momentumtwirling}
\end{align}
Here we have used the fact that the eigenspaces of the total center of mass operator $X_C=\frac{mx+MX}{m+M}$, are all two-fold degenerate, namely $\ket{X_C, \pm\frac{L}{2}}=\ket{(\frac{m}{M}+1)X_C\mp\frac{mL}{2M}}_{RF}\otimes\ket{\pm\frac{mL}{2M}}_S$~\footnote{We note that as only the relative distance between the two states matters, when computing the overlap of the two gaussian distributions.  The relative distance is of course given by $\frac{mL}{M}$.}.

A similar analysis gives for the joint system-plus-device state $\proj{\text{RF}}\otimes\Psi$
\begin{equation}
\cG[\rho_{RF}\otimes \Psi]=\bigoplus_{z=-\infty}^{\infty}\left(\begin{matrix} \frac{\mathrm{RF}(X)}{2} &0\\
				      									0&   \frac{\mathrm{RF}(X+\frac{mL}{M})}{2}\end{matrix}\right).
\label{app:momentumbasismixed}													
\end{equation}
Computing the trace distance of the states in Eqs.~(\ref{app:momentumtwirling},~\ref{app:momentumbasismixed}) one obtains
Eq.~\eqref{eq:massive size}.
\section{Macroscopic RFs and relative orientations}
\label{app:orientationtwirling}

Here we consider two copies of a macroscopic superposition for spin, Eq.~\eqref{eq:spinmacro}, where one copy serves as the required RF for the other, in order to certify macroscopic superpositions. Specifically, we are interested in the situation where the total system is composed of $2N$ spin-$\frac{1}{2}$ particles and our goal is to distinguish between the state  
\be
\ket{\Psi}\ket{\Psi} = \frac{1}{2}\left(\ket{\uparrow}^{\otimes N} + \ket{\downarrow}^{\otimes N}\right)^{\otimes 2}
\label{app:macrorf}
\ee
and the corresponding mixture
\be
\Psi\otimes\Psi=\frac{1}{4}\left((\proj{\uparrow})^{\otimes N}+(\proj{\downarrow})^{\otimes N}\right)^{\otimes 2}.
\label{app:mixed}
\ee
The difference between the two density operators $\Delta \equiv \proj{\Psi}\otimes\proj{\Psi} - \Psi\otimes\Psi$ is given by 
\begin{align} \nonumber
\Delta = \underbrace{\frac{1}{2}(\ketbra{\uparrow}{\downarrow}^{\otimes N}+\ketbra{\downarrow}{\uparrow}^{\otimes N}) \otimes \Psi}_{\equiv\Delta_1} + \\  \nonumber\underbrace{\frac{1}{2}\Psi\otimes(\ketbra{\uparrow}{\downarrow}^{\otimes N}+\ketbra{\downarrow}{\uparrow}^{\otimes N})}_{\equiv\Delta_2}+\\
\underbrace{\frac{1}{4}\left(\ketbra{\uparrow}{\downarrow}^{\otimes N}\!\!\otimes\ketbra{\downarrow}{\uparrow}^{\otimes N} +\ketbra{\downarrow}{\uparrow}^{\otimes N}\!\!\otimes\ketbra{\uparrow}{\downarrow}^{\otimes N}\right)}_{\equiv\Delta_{12}}
\label{app:Delta}
\end{align}

As the relevant symmetry group is that of orientations, i.e., $G=su(2)$, the $G$-twirling operation, acting on an arbitrary state $\varrho\in\cB(\cH^{\otimes 2N})$, is given by 
\be
\cG[\varrho]=\sum_{J=0}^N\left(\cD_{\cM_J}\otimes\cI_{\cN_J}\right)\Pi_J \varrho \Pi_J,
\label{app:su(2)twilr}
\ee
where the total state space $\cH^{\otimes 2N}$ decomposes according to Eq.~\eqref{app:reducibility} with $\cM_J=\text{span}\{\ket{J,M}\}_{M=-J}^J$ the spaces of spin-$J$ systems (on which rotations act non-trivially), and $\mathcal{N}_J=\text{span}\{\ket{\alpha_{J,k}}\}_{k=1}^{n_J}$, with $n_J=\frac{2J+1}{N+J+1}\binom{2N}{N+J}$, the spaces on which permutations of qubits act non-trivially.  Recall that the $\cD_{\cM_J}$ is the completely depolarizing map, $\cI_{\cN_J}$ is the identity map and $\Pi_J=\sum_{M=-J}^J\sum_{k=1}^{n_J}\proj{J,M,\alpha_{J,k}}$.  Observe that as the $G$-twirling map destroys coherences between states with a different number of spin-up and spin-down components, $\cG(\Delta_1)=\cG(\Delta_2)=0$. 

The measurement that maximizes the trace distance of Eq.~\eqref{eq:tracedistance} corresponds to simply measuring the total angular momentum $J$. It follows that for this measurement the difference between the probabilities $\Delta P_j \equiv P_J(\ket{\Psi}\ket{\Psi})-P_J(\Psi\otimes \Psi)$ to observe the outcome $J$ for the two states is 
\begin{align}\nonumber
\Delta P_J&=\sum_{M=-J}^J \tr\,(\Pi_{J,M} \cG[\Delta_{12}])=\sum_{M=-J}^J \tr\, (\cG[\Pi_{J,M}]\Delta_{12})\\   \nonumber
&=\sum_{M=-J}^J \tr\, (\Pi_{J,M}\Delta_{12})=\tr\, \Pi_{J,0} \Delta_{12}\\
&= n_j\sum_{k=1}^{n_J} \bra{J,0,\alpha_{J,k}}\Delta_{12}\ket{J,0,\alpha_{J,k}}.
\label{app:deltaP}
\end{align}

For ease of calculation it is preferable to use the following overcomplete basis for the expansion of $\Pi_{J,0}$ in terms of qubits.  Define
\be
\ket{J,0,\bm{k}}=\ket{\Psi^-}_{(12)}\ldots\ket{\Psi^-}_{2(N-J-1),2(N-J)-1}\ket{J,0}
\label{app:overcomplete}
\ee
where $\bm{k}=(1,2;\ldots,2(N-J-1),2(N-J)-1)$ labels the $N-J$ singlets, and $\ket{J,0}$ is a totally symmetric state of the remaining $2J$ qubits with $M=0$.  Then the projector $\Pi_{J,0}$ is proportional to the sum over all \defn{inequivalent} projectors $\ket{J,0,\bm k}$ of Eq.~\eqref{app:overcomplete} obtained by permuting all $2N$ qubits with the proportionality constant given by the dimension $n_J$~\cite{Cirac:99}. In order to compute Eq.~\eqref{app:deltaP} we need to determine which of these projectors have a non-zero overlap with $\Delta_{12}$ as well as the value of this overlap.

To avoid counting each singlet twice we impose the first qubit in each singlet to have a lower position than the second. There are in total $\binom{2N}{2N-2J}$ to chose the $2N-2J$ particles to carry the singlet states, and on top of this $(2N-2J-1)!!$ inequivalent ways to pair the particles in this group, yielding a total of $\binom{2N}{2N-2J}(2N-2J-1)!!$ inequivalent states. From the form of $\Delta_{12}$ it is clear that a non-zero overlap is achieved if and only if the first particle from each singlet can be found in any of the first $0-N$ positions (first half) and the second in any of the  $(N+1)-2N$ positions (second half).  
There in total $\binom{N}{N-J}^2$ ways to chose $N-J$ spins from the first half and $N-J$ spins from the second.  Finally there are $(N-J)!$ ways of pairing particles from the first group to those of the second
yielding that total number of state with non-zero overlap with $\Delta_{12}$ of $\binom{N}{N-J}^2(N-J)!$. Hence, the probability that a random state from this overcomplete basis overlaps with $\Delta_{12}$ is given by
\be
\beta=\frac{\binom{N}{N-J}^2(N-J)!}{\binom{2N}{2N-2J}(2N-2J-1)!!}.
\ee  
Using the fact that 
\be
\braket{\uparrow\downarrow}{\Psi^-}\!\!\braket{\Psi^-}{\downarrow\uparrow}=\braket{\downarrow\uparrow}{\Psi^-}\!\!\braket{\Psi^-}{\uparrow\downarrow}= -\frac{1}{2}
\ee
Eq.~\eqref{app:deltaP} reads
\begin{align}\nonumber
\Delta P_J&=n_J\beta\frac{(-1)^{N-J}}{2^{N-J+1}}\tr(\proj{J,0}((\ketbra{\uparrow}{\downarrow})^{\otimes J}))\\
&=n_J\beta\frac{(-1)^{N-J}}{2^{N-J+1}\binom{2J}{J}},
\end{align}
and the trace distance yields
\be
t =\frac{1}{2} \sum_{j=0}^{N} |\Delta P_J| = \frac{1}{4}.
\ee

\section{General macroscopic quantum states}
\label{app:macro_states}
We now address the demand on the size of the relevant RF needed in order to distinguish a general macroscopic quantum state 
\be
\ket{\Psi}= \sum_n \psi_n \ket{n},
\label{eq:app_general_macro}
\ee
where we assume without loss of generality $\psi_n\in\mathbb{R},\, \forall n$, from the suitable semi-classical state given by 
\be
\rho_\text{cl}(\sigma) = \cE_\sigma(\ket{\Psi}) = \sum_{n,m} \psi_n \psi_m c_{n-m} \ketbra{n}{m}
\label{eq:app_semi_clasic}
\ee 
with $c_{n-m} \equiv e^{-\frac{(n-m)^2}{2\sigma^2}}$. As mentioned in Sec.~\ref{sec:states nogo} the matrix $\rho_\text{cl}(\sigma)$ is a valid density matrix, because 
it is given by the action of a gaussian dephasing channel $\cE_\sigma$ on the state $\ket{\Psi}$ of Eq.~\eqref{eq:app_general_macro}.

As we are interested in detecting quantum coherence between far away components $\ket{n}$ and $\ket{m}$ let us introduce a size parameter $N$.  Furthermore, we shall assume a gaussian noise channel with $\sigma=N^\beta$, and consider the coherences of the semi-classical state $\rho_\text{cl}(N^\beta)$ for $n-m=N^c$.  In the limit of $N\to\infty$ these coherence terms vanish if $c>\beta$ meaning that the state  $\rho_\text{cl}(N^\beta)$  exhibits no coherence between terms separated by $N^\beta$. We will now show that, if the size of the measuring device is limited, for some choices of $\beta < 1$, and after we apply the $G$-twirling map of Eq.~\eqref{app:doublereptwirling}, the state $\rho_\text{cl}[N^\beta]$ is indistinguishable from $\ket{\Psi}$.

To that end, consider a RF whose size (given by its variance with respect to the conserved quantity in question) is $M= N^{2-\varepsilon}$. In the large $N$ limit the state of the RF is well approximated by a Gaussian, which we assume without loss of generality to be centred at zero, i.e.,
\begin{equation}
\ket{\text{RF}}= \sum_{n=-\infty}^{\infty} r_n\ket{n} \\ 
\label{app:eq_rf}
\end{equation}
After applying the $G$-twirling map of Eq.~\eqref{app:doublereptwirling} we obtain for the macroscopic state 
\begin{equation}
\cG\left[\proj{\Psi}\otimes\proj{\text{RF}}\right]=\sum_{k} P_k\proj{\Psi^{(k)}}
\label{eq:app_twirled_macro_state}
\end{equation}
where
\begin{align}\nonumber
\ket{\Psi^{(k)}}&=\sum_{n}\frac{\psi_nr_{k-n}}{\sqrt{P_k}}\ket{n}_S\ket{k-n}_{\text{RF}}\\
&\equiv\sum_n\lambda_n^{(k)}\ket{n^{(k)}},
\label{eq:app_macro_state_sectors}
\end{align}
and $P_k \equiv\sum_{n} (\psi_n r_{k-n})^2$.  Similarly the $G$-twirling of the semi-classical state with the RF reads
\begin{equation}
\cG\left[\rho_{\text{cl}}(N^\beta)\otimes\proj{\text{RF}}\right]=\sum_k P_k\rho_{\text{cl}}^{(k)}(N^\beta),
\label{eq:app_twirled_semi-classical}
\end{equation}
where
\be
\rho_{\text{cl}}^{(k)}(N^\beta)=\sum_{n, m}\lambda_n^{(k)}\lambda_m^{(k)}e^{\frac{-(n-m)^2}{2N^{2\beta}}}\ketbra{n^{(k)}}{m^{(k)}}.
\label{eq:app_semi_classical_sector}
\ee

Due to the block diagonal structure the trace distance is additive with respect to the blocks $k$
\begin{align}\nonumber
t&=\sum_k P_k \frac{1}{2}\left|\left|\proj{\Psi^{(k)}}-\rho_\text{cl}^{(k)}(N^\beta)\right|\right|\\
&\equiv\sum_k P_k  t_k
\end{align}
Using the relation $t\leq \sqrt{1-F^2}$, between the trace distance and fidelity~\cite{MikeIke} we obtain
\begin{align}\nonumber
t_k&\leq \sqrt{1 - \bra{\Psi^{(k)}} \rho^{(k)}_\text{cl}(N^\beta)\ket{\Psi^{(k)}} }\\ \nonumber
&=\sqrt{\sum_{n,m} (\lambda^{(k)}_n)^2 (\lambda^{(k)}_m)^2 \left(1-e^{-\frac{(n-m)^2}{2 N^{2\beta}}}\right)}\\ \nonumber
&\leq \sqrt{\sum_{n,m} (\lambda^{(k)}_n)^2 (\lambda^{(k)}_m)^2 \frac{(n-m)^2}{2 N^{2\beta}}}\\ \nonumber
&=\frac{1}{N^\beta}\sqrt{\sum_n n^2 (\lambda^{(k)}_n)^2  - \left(\sum_n n (\lambda^{(k)}_n)^2\right)^2}\\
&=\frac{\sqrt{\text{Var} \left((\lambda^{(k)})^2\right)}}{N^\beta},
\label{eq:app_trace_distance_blocks}
\end{align}
where we have used the inequality $1- e^{-x^2}\leq x^2$ in going from the second to the third line.  It follows that
\be
t\leq  \frac{\sum_k P_k \sqrt{ \text{Var}\left((\lambda^{(k)})^2\right)}}{N^\beta},
\label{eq:app_trace_dist}
\ee
and by Jensen's inequality, $\mathbb{E}\left(\sqrt{X_k}\right)\leq \sqrt{\mathbb{E}\left(X_k\right)}$, 
\be
t\leq  \sqrt{\frac{\sum_k P_k  \text{Var}\left((\lambda^{(k)})^2\right)}{N^{2\beta}}},
\ee
where $\text{Var}\left((\lambda^{(k)})^2\right)=\mean{n^2}_k -\mean{n}_k^2$ is the variance of the probability distribution $\{(\lambda^{(k)})^2\}$ in sector $k$.  
Employing Jensen's inequality a second time, $\sum_k P_k \mean{n}_k^2\geq (\sum_k P_k \mean{n}_k)^2$, and recalling that $(\lambda^{(k)})^2=\frac{\psi_n^2 r_{n-k}^2}{P_k}$,  we arrive at
\begin{equation}
t\leq \left[ \frac{ \sum_n n^2 (\sum_k \psi_{n-k}^2) r_{n}^2 - ( \sum_n n (\sum_k \psi_{n-k}^2) r_{n}^2)^2}{N^{2\beta}}\right]^{1/2}.
\label{eq:app_final_trace_dist}
\end{equation}
As $\sum_k \psi_{n-k}^2=1$, and recalling that $r_{n-k}^2$ has variance $N^{2-\epsilon}$, the last expression becomes
\be
t\leq \sqrt{\frac{\text{Var}\left( r_n^2\right)}{N^{2\beta}}} = \sqrt{N^{2-\varepsilon - 2\beta}},
\label{eq:app_final_trace_dist_2}
\ee
and goes to zero so long as $\beta > 1-\frac{\varepsilon}{2}$. 

Thus, we have shown that with a RF of size $N^{2-\varepsilon}$ the superposition state $\ket{\Psi}$ is indistinguishable from a semi-classical state $\rho_\text{cl}$, which does not exhibit any quantum coherence on the scale $n-m=N^c$ for
\be
c> 1-\frac{\varepsilon}{2}.
\ee
In the case where the RF state is itself a mixture $\rho_{\text{RF}}=\sum_i q_i \proj{\text{RF}_i}$, one obtains 
$t\leq \sum_i q_i \sqrt{\frac{\text{Var}\Big((r_n^{(i)})^2\Big)}{N^{2\beta}}}$ instead of Eq.~\eqref{eq:app_final_trace_dist_2}.  Another application of Jensen's inequality implies 
\begin{equation}
t\leq \sqrt{\frac{\sum_i q_i \text{Var}\Big((r_n^{(i)})^2\Big)}{N^{2\beta}}}
\end{equation}

\bibliographystyle{apsrev4-1}
\bibliography{macro}
\end{document}